%% file: NonMaxSol_forsub.tex
\documentclass[12pt]{article}

\usepackage{amssymb}
\usepackage{amsfonts}
\usepackage{amsmath}
\usepackage{graphicx}
\usepackage{amsthm}
\usepackage{amsxtra}
\usepackage{caption}
\usepackage[]{subfigure}
\usepackage{cite}
\usepackage{hyperref}
\usepackage{fullpage}

\numberwithin{equation}{section}

\include{gsmacros}

\author{Dmitry E. Pelinovsky\footnote{Department of Mathematics and
    Statistics, McMaster University, Hamilton, Ontario, Canada, L8S 4K1}
 \and
  Gideon Simpson\footnote{Department of Mathematics, University of
    Toronto, Toronto,  Ontario, Canada, M5S 2P8}
 \and
  Michael I. Weinstein\footnote{\small Department of Applied Physics
    and Applied Mathematics, Columbia University, New York, NY, USA,
    10027}
}

\title{Broad Band Solitons in a Periodic and Nonlinear Maxwell System}

\begin{document}
\maketitle

\begin{abstract}
  We consider the one-dimensional Maxwell equations with low contrast
  periodic linear refractive index and weak Kerr nonlinearity. In this
  context, wave packet initial conditions with a single carrier
  frequency excite infinitely many resonances. On large but finite
  time-scales, the coupled evolution of backward and forward waves is
  governed by nonlocal equations of resonant nonlinear geometrical
  optics. For the special class of solutions which are periodic in the
  fast phase, these equations are equivalent to an infinite system of
  nonlinear coupled mode equations, the so called {\it extended
    nonlinear coupled mode equations}, xNLCME.  Numerical studies support
  the existence of long-lived spatially localized coherent structures,
  featuring a slowly varying envelope and a train of {\it carrier
    shocks}. Thus, it is natural to study the localized coherent
  structures of xNLCME.

  In this paper we explore, by analytical, asymptotic and numerical
  methods, the existence and properties of spatially localized
  structures of the xNLCME system, which arises for a refractive index
  profile consisting of periodic array of Dirac delta functions.

  We consider, in particular, the limit of small amplitude solutions
  with frequencies near a band-edge. In this case, stationary xNLCME
  is well-approximated by an infinite system of coupled, stationary,
  nonlinear Schr\"odinger equations, the {\it extended nonlinear
    Schr\"odinger system}, xNLS.  We embed xNLS in a one-parameter
  family of equations, xNLS$^\epsilon$, which interpolates between
  infinitely many decoupled NLS equations ($\epsilon=0$) and xNLS
  ($\epsilon=1$). Using bifurcation methods we show existence of
  solutions for a range of $\epsilon\in(-\epsilon_0,\epsilon_0)$ and,
  by a numerical continuation method, establish the continuation of
  certain branches all the way to $\epsilon=1$. Finally, we perform
  time-dependent simulations of truncated xNLCME and find the
  small-amplitude near-band-edge gap solitons to be robust to both
  numerical errors and the NLS approximation.

\end{abstract}

\clearpage

%{\footnotesize{\tableofcontents}}

\section{Introduction and Overview}

Nonlinear waves in periodic structures have been a subject of great
interest for many years. Early interest arose from the possibility of
balancing the {\it band dispersion} of the periodic structure with the
nonlinearity to form soliton-like structures; see, for example,
\cite{desterke1994gs,Eggleton-Slusher} and references cited therein.
While such a heterogeneous medium possesses the same soliton-producing
ingredients of dispersion and nonlinearity as found in the well known
Korteweg--de Vries (KdV) and nonlinear Schr\"odinger (NLS) equations
which govern nonlinear dispersive waves in spatially homogeneous
media, the periodic variations of such an optical medium introduces
additional possibilities.  Indeed, changing the periodicity and
material contrasts of the medium may permit tuning of the dispersive
properties, {\it e.g.} the length scale on which a soliton can form may be
altered. Thus, nonlinear and periodic structures are natural candidates for device design and
applications. An example is the formation of centimeter-scale {\it gap
  solitons} in periodic optical fiber gratings. Such states have been
shown to propagate at a fraction of the speed of light and have been
proposed in schemes for optical storage and buffering; see, for
example, \cite{GSW}. \medskip

In the simplest setting, nonlinear electromagnetic waves in a
one-dimensional periodic structure are governed by a nonlinear Maxwell
equation:
\begin{equation}
  \label{eq:maxwell1}
  \partial^2_t \paren{n^2(z) E + \chi \abs{E}^2E} = \partial^2_z E.
\end{equation}
Here, $\chi>0$ is the Kerr nonlinearity coefficient,
\cite{boyd-optics}. We assume a low-contrast, periodic refractive
index profile, $n(z)$, with mean $n_0$, given by
\begin{equation}\label{index}
  n(z) = n_0 + \eps N(z),\quad n_0 > 0,\ \  N(z) = N(z+2\pi), \ \ 0<\eps\ll1;
\end{equation}
$n(z)$ is real-valued; no energy-dissipation has been included.  The
fluctuating part of the refractive index, $N(z)$, can be expanded in
the Fourier series
\begin{equation}
  \label{Fourier-series-N}
  N(z) \equiv \sum_{p \in \mathbb{Z}} N_p e^{\mathrm{i} p
    z}, \quad N_{-p} = \bar{N}_p, \;\; p \in \mathbb{Z}.
\end{equation}

For simplicity, let us assume $N_2\ne0$. Then strong dispersion is
excited by initial conditions of wave-packet type, {\it i.e.} a slowly
modulated plane wave of a single frequency, chosen to be in (Bragg)
resonance with the $\pi$-periodicity of the medium:
\begin{equation}
  E(z,t=0) = \eps^{1\over 2} \left[ E^+_1(\eps z, 0) e^{\mathrm{i} z}
    + E^-_1(\eps z, 0)e^{-\mathrm{i}z} + \cc  \right],
  \label{monochrom-data}\end{equation}
where $E_1^{\pm}(Z,0)$ are spatially localized in $Z=\eps z$.  This
resonance strongly couples backward and forward propagating waves.  In
the choice of initial condition \eqref{monochrom-data}, dispersive
effects which are set by the medium contrast, of size
$\bigo(\epsilon)$, have been balanced with nonlinear effects, by
choosing the amplitude to be of size
$\bigo(\epsilon^{1\over2})$.\bigskip

Suppose we make a formal multiple scale expansion based on the ansatz:
\begin{gather}
  \label{eq:nlcme_ansatz}
  E(z,t) = \eps^{1\over 2} \left[ E^+_1(Z,T) e^{\im (z- v_g t)}
    + E^-_1(Z,T)e^{-\im(z+v_gt)} + \cc  + \bigo(\eps) \right],\\
  T=\eps t, \quad Z=\eps z,\quad v_g\equiv 1/n_0\nn
\end{gather}
Then if we only account for the principal harmonics, we shall arrive at
the nonlinear coupled mode equations (NLCME) for $E^\pm_1(Z,T)$:
\begin{subequations} \label{eq:nlcme}
  \begin{align}
    \dT E^+_1 + v_g \dZ E^+_1 &= \im v_g^2 \paren{N_0 E_1^+ + N_2 E_1^-}+\im \Gamma\paren{\abs{E_1^+}^2 +  2 \abs{E_1^-}^2 } E_1^+,\\
    \dT E^-_1 - v_g \dZ E^-_1 &= \im v_g^2 \paren{ \bar{N}_2 E_1^+ +
      N_0 E_1^-}+ \im \Gamma \paren{\abs{E_1^-}^2 + 2 \abs{E_1^+}^2 }
    E_1^-,
  \end{align}
\end{subequations}
where $\Gamma \equiv 3 \chi/(2n_0^3)$.  $E^\pm_1$ denote slowly
varying forward and backward wave amplitudes; see
\cite{desterke1994gs} and references cited therein for details.

NLCME has been rigorously derived as a leading order model in numerous
contexts.  For one-dimensional propagation of electromagnetic waves in
nonlinear and periodic media, it was rigorously derived from the
anharmonic Maxwell-Lorenz model in \cite{goodman01npl}.  Derivations
from the Klein-Fock as well as Gross-Pitaevskii equations have also
been obtained; see
\cite{schneider2001ncm,schneider2003eas,pelinovsky2007jcm,pelinovsky2008mgs}.
Explicit localized stationary solutions, called {\it gap solitons},
for NLCME are given in \cite{aceves1989sit,christodoulides-joseph} The
linear stability of the gap solitons was studied in \cite{ChPel}, and
a linear, multi-dimensional, analog of NLCME was examined in
\cite{agueev2005mwr}.

However, NLCME is {\it not} the correct mathematical description of
weakly nonlinear and weakly dispersive waves in the nonlinear and
periodic Maxwell equation \eqref{eq:maxwell1}, \eqref{index}. The
deficiency of the NLCME system, \eqref{eq:nlcme}, stems from the
unperturbed ($\epsilon=0$) equation being the {\it non-dispersive}
one-dimensional wave equation. Due to nonlinearity, a single frequency
initial condition, \eqref{monochrom-data}, excites infinitely many
resonances, since $e^{\im m (z\pm t/n_0)}, m\in\mathbb{Z}$ all lie in
the kernel of the unperturbed operator,
$n_0^2\partial_t^2-\partial_z^2$.  In contrast, other models, such as
the aforementioned anharmonic Maxwell-Lorenz system and the
Gross-Pitaevskii equation, remain dispersive in the $\eps=0$ limit;
this precludes infinitely many resonant modes.

In \cite{simpson2010}, nonlocal equations derived from nonlinear
geometrical optics and an equivalent system of infinitely many coupled
PDEs, which take into account the infinitely many resonances, were
systematically studied.  One begins with the general weakly nonlinear
ansatz,
\begin{equation}
  E(z,t) = \eps^{1\over 2} \left[ E^+(Z,T,z- v_g t) + E^-(Z,T,z+ v_g
    t) + \bigo(\eps) \right],\label{Ansatz}
\end{equation}
which need not be nearly monochromatic.  A necessary condition for the
error term in \eqref{Ansatz} to be of order $\epsilon$ on the time
interval $0\le t\le \bigo\left(\eps^{-1}\right)$ is that the forward
and backward wave components, $E^\pm(Z,T,\phi_\pm)$, $\phi_\pm = z \mp
v_g t$, satisfy the system of nonlocal evolution equations:
\begin{subequations}\label{nlgo}
  \begin{gather}
    \label{eq:Ep-nlgo}
    \begin{split}
      (\partial_T + v_g \partial_Z + v_g^2 N_0 \partial_{\phi}) E^+ &=
      v_g^2 \partial_\phi \left[ \frac{1}{2\pi} \int_{-\pi}^{\pi}
        N(\phi + \theta) E^-(Z,T,\phi + 2 \theta) d\theta \right] \\
      &\quad +\frac{\Gamma}{3} \partial_{\phi} \left[ (E^+)^3 + 3
        \left( \frac{1}{2\pi} \int_{-\pi}^{\pi} |E^-(Z,T,\theta)|^2
          d\theta \right) E^+ \right],
    \end{split}
    \\
    \label{eq:Em-nlgo}
    \begin{split}
      (\partial_T - v_g \partial_Z - v_g^2 N_0 \partial_{\phi})
      E_p^-&= -v_g^2 \partial_\phi\left[ \frac{1}{2\pi}
        \int_{-\pi}^{\pi}
        N(\phi - \theta) E^+(Z,T,\phi - 2 \theta) d \theta \right] \\
      &\quad - \frac{\Gamma}{3} \partial_{\phi}\left[ (E^-)^3 + 3
        \left( \frac{1}{2\pi} \int_{-\pi}^{\pi} |E^+(Z,T,\theta)|^2
          d\theta \right) E^- \right].
    \end{split}
  \end{gather}
\end{subequations}
While we have omitted the $\pm$ subscripts on $\partial_\phi$
derivatives for the sake of brevity, the reader should note that in
recovering the primitive field, as in \eqref{Ansatz}, $E^+$ must be
evaluated at $\phi_+$ and $E^-$ must be evaluated at
$\phi_-$. $E^\pm(Z,T, \phi_\pm)$ are assumed to be $2\pi$-periodic in
their $\phi_\pm$ arguments. A similar, but more general system of
integro-differential equations was obtained in 
\cite{simpson2010}, though in that work, the authors set $N_0 = 0$ and $v_g = 1$.

If we expand $E^\pm(Z,T,\phi)$ in a Fourier series with respect to the
phase variable $\phi$,
\begin{equation}
  \label{Fourier-series-E-leading}
  E^{\pm}(Z,T,\phi) = \sum_{p \in \mathbb{Z}} E^{\pm}_p(Z,T) e^{\pm \mathrm{i} p \phi},
\end{equation}
the nonlocal system \eqref{nlgo} may be re-expressed as a system of
{\it infinitely} many nonlinear coupled mode (differential) equations
for the Fourier mode coefficients, indexed by $p \in \Z$:
\begin{subequations}\label{e:mode_intro}
  \begin{gather}
    \label{eq:mode_intro_p}
    \begin{split}
      \dT \Ep_p +v_g \dZ \Ep_p &= \mathrm{i}p v_g^2 (N_0 E_p^+ + N_{2p}{\Em_{p}})  \\
      &\quad+ \mathrm{i}p\frac{\Gamma}{3} \bracket{ \sum_{q,r \in
          \mathbb{Z}} \Ep_q \Ep_r {\bar{E}^+_{q+r-p}} +
        3\paren{\sum_{q \in \mathbb{Z}} \abs{\Em_q}^2} \Ep_p },
    \end{split}
    \\
    \label{eq:mode_intro_m}
    \begin{split}
      \dT \Em_p - v_g \dZ \Em_p &= \mathrm{i}p v_g^2 (N_{-2p}{\Ep_{p}} + N_0 E_p^-) \\
      &\quad +\mathrm{i}p\frac{\Gamma}{3} \bracket{\sum_{q,r \in
          \mathbb{Z}} \Em_q \Em_r {\bar{E}^-_{q+r-p}} +
        3\paren{\sum_{q \in \Z} \abs{\Ep_q}^2} \Em_p }.
    \end{split}
  \end{gather}
\end{subequations}
In \cite{simpson2010} the infinite system of PDEs \eqref{e:mode_intro}
is referred to as the {\it extended nonlinear coupled mode equations}
or xNLCME. Thus xNLCME is an extension of the classical NLCME
\eqref{eq:nlcme}, appropriate for {\it highly resonant} settings, such
as the weakly periodic and nonlinear Maxwell model
\eqref{eq:maxwell1}. Truncation of xNLCME to a single mode,
$E^\pm_1(Z,T)$, yields NLCME, \eqref{eq:nlcme}, which, as noted, has
spatially localized gap-soliton solutions.

Numerical simulations of the primitive nonlinear and periodic
Maxwell's equations, \eqref{eq:maxwell1}, give evidence of two
phenomena.  First, there appear to be long-lived spatially localized
coherent structures.  Second, within such spatially localized
structures, a train of {\it carrier shocks} can form.  These
structures appear to be well described by xNLCME, \cite{simpson2010}.

The nonlinear Maxwell equation, \eqref{eq:maxwell1}, does not
incorporate any effects of chromatic dispersion which, as in the
anharmonic Maxwell-Lorentz model \cite{goodman01npl}, {\it takes off
  resonance} the higher harmonics.  However, chromatic dispersion on
the length scales of many experiments is a negligible effect,
\cite{ess1996}.  Moreover, there are experimentally realizable regimes
in which pulses with spectral content near the zero dispersion point
are propagated \cite{ranka2000}. In these experiments, a broad band
{\it super continuum} is generated. The carrier shocking mentioned
above is a possible source of such broad band emission.

\bigskip

In this paper, we explore, by analytical, asymptotic and numerical
methods, the existence and properties of spatially localized
structures of xNLCME.  These coherent solutions have a full spectrum
of active temporal frequencies and we therefore refer to them as {\it
  broad band solitons}. An earlier step in this direction was taken in
\cite{Tasgal:2005p6335}, where the authors studied what amounts to a
truncation of xNLCME to first and third harmonics. Studying the
problem numerically, they found evidence for spatially localized
solutions that they called {\it polychromatic} solitons.  \bigskip

We focus on the stationary, small amplitude, near band edge,
approximation of xNLCME for a particular refractive index consisting
of an infinite periodic array of Dirac delta functions.  In this
regime, xNLCME is well-approximated by an infinite system of coupled
nonlinear Schr\"odinger equations, the {\it extended nonlinear
  Schr\"odinger system}, xNLS.  We embed xNLS in a one-parameter
family of equations, xNLS$^\epsilon$, which continuously interpolates
between a system of infinitely many decoupled NLS equations
($\epsilon=0$) and xNLS ($\epsilon=1$). Using bifurcation methods,
based on the Lyapunov-Schmidt method and the implicit function
theorem, we prove the existence of solutions for a range of
$\epsilon\in(-\epsilon_0,\epsilon_0)$.  By numerical continuation
method, we establish the persistence of \underline{certain} branches all the way
to $\epsilon=1$ for finite truncations of xNLS$^\epsilon$. Finally, we
perform time-dependent simulations of xNLCME and find the small
amplitude near band edge gap solitons to be robust.

\bigskip

{\bf Outline of the paper:} In Section \ref{sec:NLCME}, we present a
direct derivation of xNLCME in the case of a periodic medium and show
the sense in which xNLCME is an infinite dimensional Hamiltonian
system. In Section \ref{sec:Gap} we heuristically determine conditions
on $N(z)$ for which we may expect exponentially localized gap
solitons. This motivates us to focus on the case where $N(z)$ is a
periodic array of delta functions.

In the small amplitude, near band edge, limit where xNLCME reduces to
xNLS, we conjecture that localized stationary solutions of xNLS exist.
Subject to this assumption, we prove in Theorem \ref{theorem-main-2}
that the gap soliton persists within xNLCME, in the asymptotic limit.
Since the energy of xNLS is bounded below, it is natural to ask where
a ground state of xNLS can be constructed
variationally. Unfortunately, standard methods to not apply due to a
loss of compactness, illustrated in Section \ref{s:proof}.  The
existence of nontrivial critical points is an open problem.

We therefore seek to construct localized states via a continuation
method.  First, we embed xNLS in a one-parameter family of systems,
xNLS$^\eps$, with $\eps=0$ corresponding to an infinite system of
decoupled NLS equations and $\eps=1$ corresponding to xNLS, the system
of interest.  In Theorem \ref{thm:xnls_mono}, we prove the existence
of gap solitons for xNLS$^\eps$ for an open interval of
$|\eps|<\eps_0$ about $\eps=0$.

We next attempt to numerically continue xNLS$^\eps$ solitons on the
interval $[0,1]$.  In order to implement the numerical continuation,
we seek approximate critical points of the xNLS$^\eps$ variational
problem.  To motivate this, in Section \ref{s:var_approx}, we replace
the variational characterization of xNLS$^\eps$ solitons by a finite
dimensional minimization problem over families of Gaussian trial
functions.  We find critical points, with sign alternating amplitudes,
of such finite dimensional approximations and give convincing
numerical evidence that some can be continued to $\eps=1$.

In Section \ref{s:gap_solitons} we compute soliton solutions of truncated
xNLS using information gleaned from the trial function
approximations, and show that they are robust in time-dependent
simulations of truncated.  Section
\ref{s:open} summarizes our findings and highlights open problems.

\bigskip

{\bf Acknowledgements:} DP and GS were supported by NSERC. MIW was
supported in part by US-NSF grants DMS-07-07850 and DMS-10-08855.

\section{Coupled Mode Equations}\label{sec:NLCME}

In Section \ref{fourier-derive}, we present a derivation of xNLCME
from Maxwell's equations using Fourier expansions of
$E^\pm(Z,T,\phi_\pm)$, in the case where $E^\pm(Z,T,\phi_\pm)$  are periodic
in $\phi_\pm$.
In Section \ref{xnlcme-properties}, we demonstrate { that} xNLCME is
an infinite dimensional Hamiltonian system with two conserved
quantities.

\subsection{Derivation of xNLCME in a Periodically Varying
  Medium}\label{fourier-derive}

For simplicity and without loss of generality, we set $n_0 = 1$ { so
  that $v_g \equiv 1$}.  We rewrite the nonlinear Maxwell equation
\eqref{eq:maxwell1} with refractive index \eqref{index} as
\begin{align}
  \label{eq:maxwell2}
  \partial^2_z E - \partial^2_t E = 2 \eps N(z) \partial^2_t E +
  \eps^2 N(z)^2 \partial^2_t E + \chi \partial^2_t \abs{E}^2 E,
\end{align}
For $\epsilon = 0$ { and $\chi = 0$}, \eqref{eq:maxwell2} simplifies
to the one dimensional wave equation with a solution, given by the
arbitrary superposition of right and left traveling waves,
\begin{equation}
  \label{e:leading_order}
  E^{(0)}(z,t) = E^+(z-t) + E^-(z+t).
\end{equation}
For $\eps$ small, we seek $E=E^\eps(z,t)$ in the form of a multiple
scale expansion
\begin{equation}
  \label{eq:expansion_ansatz}
  E(z,t) = \eps^{1\over 2} \left( E^{(0)}(Z,T;z,t) + \eps E^{(1)}(Z,T;z,t) + \bigo(\eps^2) \right),
\end{equation}
where $Z = \eps z$ and $T = \eps t$ are slow spatial and temporal
scales. Substituting \eqref{eq:expansion_ansatz} into
\eqref{eq:maxwell2}, we obtain at first order in $\eps$:
\begin{equation}
  \label{eq:order1}
  \left( \partial_z^2 - \partial_t^2 \right) E^{(1)} = 2 \left( \partial_t \partial_T - \partial_z \partial_Z \right) E^{(0)} + 2 N(z) \partial_t^2 E^{(0)} + \chi \partial_t^2 \abs{E^{(0)}}^2 E^{(0)}.
\end{equation}
The right-hand-side of \eqref{eq:order1} generates resonant terms
along the characteristics of the wave equation, leading to secular
growth of the correction $E^{(1)}$ in $(z,t)$.  The slow evolution in
$(Z,T)$ is determined to remove these secular terms.

We begin by expanding $E$ in a Fourier series:
\begin{equation}
  \label{Fourier-series-E}
  E^{\pm}(Z,T;z,t) = \sum_{p \in \mathbb{Z}} E^{\pm}_p(Z,T) e^{\pm\mathrm{i} p (z \mp t)}, \quad
  E^{(1)}(Z,T;z,t) = \sum_{p \in \mathbb{Z}} E^{(1)}_p(Z,T;t) e^{\mathrm{i} p z}.
\end{equation}
Since $E^\pm$ are real-valued,
\begin{equation}
  \label{eq:conjugate}
  \bar{E}^\pm_p(Z,T) = E^\pm_{-p}(Z,T), \quad p \in \mathbb{Z}.
\end{equation}

Substituting \eqref{Fourier-series-E} into \eqref{eq:order1}, the
terms of the equation proportional to $e^{\im pz}$ are:
\begin{align*}
  \paren{\partial_t^2 + p^2}E_p &= 2\im p (\dT + \dZ) E_p^+ e^{-\im
    pt}
  - 2\im p (\dT - \dZ) E_{-p}^- e^{\im pt}\\
  &\quad +2\sum_q q^2\paren{N_{p-q} E_{q}^+ e^{-\im qt} + N_{p+q}
    E_{q}^-
    e^{-\im q t}}\\
  &\quad + \chi \sum_{q,r} p^2 E_q^+ E_r^+ \bar E_{q + r-p}^+ e^{-\im
    pt} + 2 \chi \sum_{q,r} (p-2q + 2r)^2E_q^+ \bar E_r^+ E_{q-r-p}^-
  e^{\im (p-2q +
    2r)t}\\
  &\quad + \chi \sum_{q,r} (p+ 2q + 2r)^2 E_q^- E_r^- \bar
  E_{-p-q-r}^+
  e^{-\im (p + 2q + 2r)t} \\
  &\quad + \chi \sum_{q,r} (p-2q-2r)^2E_q^+ E_r^+ \bar E_{p-q-r}^-
  e^{\im (p-2q-2r)t}\\
  &\quad +2 \chi \sum_{q,r} (p + 2q - 2 r)^2 E_q^- \bar E_r^-
  E_{p+q-r}^+ e^{-\im (p + 2q - 2r)t} + \chi \sum_{q,r} p^2 E_{q}^-
  \bar E_{r}^- E_{-p-q+r}^- e^{\im pt},
\end{align*}
where all sums are taken over $\Z$.  Removing the terms resonant with $e^{\im pt}$, we obtain
\begin{subequations}
  \label{e:E_mode}
  \begin{equation}
    \label{eq:Ep_mode}
    \begin{split}
      (\dT + \dZ) E_p^+ = & \im p\paren{N_0 E_p^+ + N_{2p} E_p^-} \\
      & + \im p\frac{\Gamma}{3} \left[ \sum_{q,r} E_q^+ E_{r}^+ \bar
        E_{q+r-p}^+ + 2
        E_0^- \sum_q E_q^+ \bar E_{q-p}^+  \right.\\
      &\left. \quad + \sum_q E_q^- E_{-q}^- \bar E_{-p}^+ + \bar
        E_0^-\sum_q E_q^+ E_{p-q}^+ + 2\sum_{q} \abs{E_q^-}^2
        E_p^+\right].
    \end{split}
  \end{equation}
  Removing terms resonant with $e^{-\im pt}$, we obtain
  \begin{equation}
    \label{eq:Em_mode}
    \begin{split}
      -(\dT -\dZ) E_{-p}^- = & \im p \paren{N_{2p} E_{-p}^+ + N_0
        E_{-p}^-}\\
      &+ \im p \frac{\Gamma}{3}\left[ \sum_{q,r} E_q^- E_r^- \bar
        E_{q+r+p}^- + 2 E_0^+\sum_q E_q \bar
        E_{p+q}^-\right.\\
      &\left. \quad+ \sum_q E_q^+ E_{-q}^+ \bar E_p^-+ \bar E_0^+
        \sum_q E_q \bar E_{-p-q}^+ +2 \sum_q
        \abs{E_{q}^+}^2E_{-p}^-\right]
    \end{split}
  \end{equation}
\end{subequations}
where we have set $\Gamma \equiv 3\chi/2$ to be consistent with
previous work, \cite{goodman01npl, simpson2010}.  Exchanging $p$ for
$-p$ in \eqref{eq:Em_mode}, we have
\[
\begin{split}
  (\dT -\dZ) E_{p}^- = & \im p \paren{N_{-2p} E_{p}^+ + N_0
    E_{p}^-}\\
  & +\im p \frac{\Gamma}{3}\left[ \sum_{q,r} E_q^- E_r^- \bar
    E_{q+r-p}^- + 2 E_0^+\sum_q E_q \bar
    E_{-p+q}^-\right.\\
  &\left. \quad+ \sum_q E_q^+ E_{-q}^+ \bar E_{-p}^-+ \bar E_0^+
    \sum_q E_q \bar E_{p-q}^+ +2 \sum_q \abs{E_{q}^+}^2E_{p}^-\right]
\end{split}
\]

At $p=0$, \eqref{e:E_mode} can be satisfied by choosing arbitrary
functions $E_0^{\pm} = E_0^{\pm}(Z \mp T)$. For simplicity, we set
$E_0^{\pm}(Z\mp T) \equiv 0$.  If we additionally invoke complex
conjugate relationship \eqref{eq:conjugate}, \eqref{e:E_mode} simplify
to xNLCME, \eqref{e:mode_intro}, from the introduction, provided $v_g
= 1$.

Finally, the nonlocal system \eqref{nlgo} can be recovered by
introducing the identities
\begin{equation}
  \label{Fourier-series-A}
  E^\pm(Z,T,\phi) = \sum_{p \in \Z} E^\pm_p(Z,T) e^{\pm \im  p \phi}, \quad
  E^\pm_p(Z,T) = \frac{1}{2\pi} \int_{-\pi}^{\pi} E^\pm(Z,T,\phi) e^{\mp\im p \phi} d \phi.
\end{equation}
Constraints \eqref{eq:conjugate} imply that $E^{\pm}$ are
real-valued.  Note that in the context of the primitive electric field
variables, $E^\pm(Z,T, \phi_\pm)$ must be evaluated at different
phases, $\phi_\pm = z \mp t$.

% We have introduced the phases, $\phi_\pm = z\mp t$ to remind the
% reader that should we wish to return to the primitive electric field
% variable as in \eqref{e:leading_order} and
% \eqref{eq:expansion_ansatz}, $E^\pm$ must be evaluated at different
% values of the fast, phase variable.

\subsection{Hamiltonian Structure of xNLCME}\label{xnlcme-properties}

Let $E^\pm_0$, and define  $H =
\int_{\mathbb{R}} \mathcal{H} dZ$, where 
\begin{equation}
  \begin{split}
    \mathcal{H} &= \frac{\mathrm{i}}{2} \sum_{p} \frac{1}{p}
    \paren{\Ep_{p} \dZ \bar{E}^+_ {p} - \Em_{p} \dZ\bar{E}^-_{p}
      - \bar{E}^+_{p} \dZ \Ep_ {p} + \bar{E}^-_{p} \dZ \Em_{p}} \\
    &\quad - N_0 \sum_{p} (|E^+_p|^2 + |E^-_p|^2) - \sum_{p} N_{2p}
    (\bar{E}^-_{-p} E^+_{-p} + E_{p}^- \bar{E}^+_{p}) \\
    &\quad - \frac{\Gamma}{6}
    \paren{\sum_{p} \bar{E}_p^+ \bar{E}_{-p}^+ } \paren{\sum_{p} E_p^-
      E_{-p}^-} - \frac{\Gamma}{6}
    \paren{\sum_{p} \bar{E}_p^- \bar{E}_{-p}^-
    } \paren{\sum_{p} E_p^+ E_{-p}^+} \\
    &\quad - \frac{\Gamma}{6} \sum_{p,q,r} \left( \bar{E}^+_{p}
      \Ep_{q}\Ep_{r} \bar{E}^+_{q+r-p} + \bar{E}^-_{p} \Em_{q} \Em_{r}
      \bar{E}^-_{q+r-p} \right)  - \frac{2
      \Gamma}{3}
    \paren{\sum_{p} \abs{\Ep_{p}}^2} \paren{\sum_{p} \abs{\Em_{p}}^2},
  \end{split}\label{hamiltonian}
\end{equation}
with all sums are over $\mathbb{Z}\setminus\{0\}$.  Then, xNLCME has
the structure of an infinite dimensional Hamiltonian system:
\begin{align}
  \dT \Ep_p = - \im p \frac{\delta H}{\delta \bar{E}^+_p},
  \quad \dT \Em_p = -\im p \frac{\delta H}{\delta\bar{E}^-_p},
  \quad p \in \mathbb{Z}\setminus\{0\}.
\end{align}

Formally, the Hamiltonian \eqref{hamiltonian} is conserved under the
flow of xNLCME. Besides the Hamiltonian, the total power $N =
\int_{\mathbb{R}} \mathcal{N} dZ$ is invariant, where the density is
\begin{equation}
  \mathcal{N} = \sum_{p \in \mathbb{Z}} \left( \abs{\Ep_{p}}^2 + \abs{\Em_{p}}^2 \right).
\end{equation}
This follows by direct computation.

Since $N_{2p} = \bar{N}_{-2p}$, $p \in \mathbb{Z}$, the symmetry of
equations \eqref{e:mode_intro} implies that if the constraint
$\bar{E}_p^\pm=E^\pm_{-p}$, associated with real initial conditions
for $E^{\pm}$, is satisfied at $T=0$, then it is satisfied for all
$T$.  Additionally, if $E^{\pm}_p$ are zero initially for even $p$,
they remain zero for all time. This allows us to restrict
\eqref{e:mode_intro} to the odd harmonics, $p \in \mathbb{Z}_{\rm
  odd}$, and set
\begin{equation}
  \label{constraints-even}
  E^{\pm}_p = 0,  \quad p \in \mathbb{Z}_{\rm even}.
\end{equation}
Under these constraints, the conserved integral \eqref{hamiltonian}
reduces to the Hamiltonian:
\begin{equation}
  \label{e:constrained_hamiltonian}
  \begin{split}
\mathcal{H} &= \frac{\mathrm{i}}{2} \sum_{p \in
      \mathbb{Z}_{\rm odd}} \frac{1}{p}
    \paren{\Ep_{p} \dZ \bar{E}^+_ {p} - \Em_{p} \dZ\bar{E}^-_{p}
      - \bar{E}^+_{p} \dZ \Ep_ {p} + \bar{E}^-_{p} \dZ \Em_{p}} \\
    &\quad - N_0 \sum_{p \in \mathbb{Z}_{\rm odd}} (|E^+_p|^2 +
    |E^-_p|^2) - 2 \sum_{p \in \mathbb{Z}_{\rm odd}} N_{2p} {E}^-_{p}
    \bar{E}^+_{p}\\
&\quad - \Gamma
    \paren{\sum_{p \in \mathbb{Z}_{\rm odd}}
      \abs{\Ep_{p}}^2} \paren{\sum_{p \in
        \mathbb{Z}_{\rm odd}} \abs{\Em_{p}}^2} \\
    &\quad - \frac{\Gamma}{6} \sum_{p,q,r \in \mathbb{Z}_{\rm
        odd}}\left( \bar{E}^+_{p} \Ep_{q}\Ep_{r} \bar{E}^+_{q+r-p} +
      \bar{E}^-_{p} \Em_{q} \Em_{r} \bar{E}^-_{q+r-p} \right).
  \end{split}
\end{equation}
As in the case of standard NLCME,
\eqref{e:constrained_hamiltonian} is unbounded from above and below
subject to the constraint of fixed $\calN$.  Thus, critical points
are expected to be of {\it infinite index}.  This suggests that
variational methods will be of limited applicability for studying the
stability of localized stationary states of xNLCME.

% sign indefinite about the zero
% solution.  This limits the applicability of the functional
% in the analysis of localized stationary solutions of xNLCME.

\section{Gap Solitons}\label{sec:Gap}

We now begin to explore the existence of localized stationary
states of xNLCME \eqref{e:mode_intro},  called {\it gap solitons}.  Setting $v_g = 1$ for
convenience, we seek solutions of the form
\begin{equation}
  \label{modes}
  E^+_p(Z,T) = e^{\im p(N_0-\Omega) T} A_p(Z), \quad
  E^-_{p}(Z,T) = e^{\im p(N_0-\Omega) T} B_p(Z), \quad p \in \mathbb{Z},
\end{equation}
where $\Omega$ is a real frequency parameter and $\{ A_p(Z),B_p(Z)
\}_{p \in \mathbb{Z}}$ are complex-valued amplitudes. Using
constraints \eqref{eq:conjugate} and \eqref{constraints-even}, we
assume
\begin{equation}
  \label{reduction}
  A_p = \bar{A}_{-p}, \quad B_p = \bar{B}_{-p}, \quad p \in \mathbb{Z}_{\rm odd}, \quad
  A_p = B_p = 0, \quad p \in \mathbb{Z}_{\rm even}.
\end{equation}
The infinite family of amplitudes $\{ A_p,B_p\}_{p \in \Z_{\rm odd}}$
satisfies the extended system of stationary equations
\begin{subequations}
  \label{stat-cme-system}
  \begin{align}
    \im A_p'(Z) + p \Omega A_p + p N_{2p} B_p + p
    \frac{\Gamma}{3} \left( 3 A_p \sum_{q \in\Zodd} |B_q|^2 +
      \sum_{q,r \in \Zodd} A_q A_r A_{p-q-r} \right) &= 0, \\
    - \im B_p'(Z) + p \Omega B_p + p \bar{N}_{2p} A_p + p
    \frac{\Gamma}{3} \left( 3 B_p \sum_{q \in\Zodd} |A_q|^2 + \sum_{q,r
        \in \Zodd} B_q B_r B_{p-q-r} \right) &= 0,
  \end{align}
\end{subequations}
with constraints \eqref{reduction}.  Linearizing about the zero
solution yields decoupled systems of differential equations with
solutions
\begin{equation}
  \label{exp-decay}
  \left[ \begin{array}{c} A_p \\ B_p \end{array} \right] \sim e^{\pm Z \sqrt{|pN_p|^2 - (p\Omega)^2}}, \quad
  p \in \Zodd.
\end{equation}
A sufficient condition for spatial localization near the zero solution
is only possible if $|\Omega| < \Omega_0 \equiv {\rm min}_{p \in \Zodd}
|N_{2p}|$, implying three possibilities:

\begin{description}
\item[Case 1, $\Omega_0 > 0$:] An example would be $N_{2p} = 1$, $p \in
  \Z$, in which case the refractive index, $N(z)$, is a periodic sequence of
  Dirac delta-functions.

\item[Case 2, $\Omega_0 = 0$ and $\min_{p \in \Zodd} |p N_{2p}| > 0$:]
  An example would be $N_{2p} = p^{-1}$, $p \in \Zodd$, for which
  $N(z)$ would correspond to a periodic sequence of step functions.

\item[Case 3, $\Omega_0 = 0$ and $\min_{p \in \Zodd} |p^2 N_{2p}|
  <\infty$:] In this case, $N(z)$ is continuous.
\end{description}

If $N_{2p} = 1$, $p \in \Zodd$, the band gap of each mode is opened, and
the widths of the band gaps grow as $\abs{p}\to \infty$.  However, because of the
coupling between the Fourier modes with amplitudes $\{A_p,B_p\}_{p \in
  \Zodd}$, the stationary localized mode (gap soliton) may only reside in
the gap of a fixed width, $|\Omega| < \Omega_0 \equiv 1$.

If $N_{2p} = \bigo(\abs{p}^{-1})$, the band gap of each mode is again
opened, but the widths are nearly constant as $\abs{p} \to \infty$.
However, the band gap for the coupled gap soliton shrinks now to zero
and the parameter $\Omega$ must be set to $0$.

If $N_{2p} = \bigo(\abs{p}^{-2})$, the widths of the band gaps shrink with
the larger values of $p$, and the exponential decay \eqref{exp-decay}
ceases as $\abs{p} \to \infty$, even if $\Omega = 0$. We do not
anticipate the existence of gap solitons in this case.

We now restrict our attention to Case 1: $\Omega_0 > 0$ and set
$N_{2p} = 1$ for all $p\in\Zodd$.  System \eqref{stat-cme-system} can now
be rewritten as an equivalent integro-differential equation:
\begin{subequations}\label{e:stat_fourier}
  \begin{align}
    \label{eq:stat_fourier}
    \begin{split}
      (-\partial_Z + \Omega \partial_{\phi}) A + \partial_{\phi} B +
      \frac{\Gamma}{3} \partial_{\phi} \left[ A^3 + 3 \left(
          \frac{1}{2\pi} \int_{-\pi}^{\pi} |B(Z,s)|^2 d s \right) A
      \right] &= 0,
    \end{split}
    \\
    \label{eq:stat_fourier}
    \begin{split}
      (\partial_Z + \Omega \partial_{\phi}) B + \partial_{\phi} A +
      \frac{\Gamma}{3} \partial_{\phi} \left[ B^3 + 3 \left(
          \frac{1}{2\pi} \int_{-\pi}^{\pi} |A(Z,s)|^2 d s \right) B
      \right] &= 0,
    \end{split}
  \end{align}
\end{subequations}
where we have introduced the two Fourier series,
\begin{equation}
  \label{Fourier-series-B}
  A(Z,\phi) = \sum_{p \in \Z_{\rm odd}} A_p(Z) e^{\im p \phi}, \quad
  B(Z,\phi) = \sum_{p \in \Z_{\rm odd}} B_p(Z) e^{\im p \phi}.
\end{equation}
We note that if one wishes to compute the primitive electric field
induced by these envelopes, care must be taken in where the phase
variable, $\phi$, is evaluated.  Indeed, the electric field
associated with $\{A_p, B_p\}_{p \in \Z_{\rm odd}}$ is given by
\begin{equation}
  \label{e:primitive_efield}
  \begin{split}
    E(z,t) &= \eps^{1\over 2} \bracket{\sum_{p\in \Zodd} e^{ \im p
        (N_0 - \Omega)\eps t}e^{\im p(z -t)} A_p(\eps z) +\sum_{p\in
        \Zodd} e^{\im p (N_0 - \Omega)\eps
        t}e^{-\im p(z+t)} B_p(\eps z) + \bigo(\eps)}\\
    & = \eps^{1\over 2}\bracket{A(\eps z, (N_0 - \Omega) \eps t + z-t)
      + B(\eps z, (N_0 -\Omega) \eps t - (z+t)) + \bigo(\eps)},
  \end{split}
\end{equation}
in agreement with the ansatz \eqref{Ansatz}.

\subsection{NLCME Gap Solitons}
As noted, the truncation of xNLCME to $E^\pm_1$ yields the classical
NLCME. We now review the details of the NLCME gap soliton.

The spatial profiles of NLCME's gap soliton are given by solutions of
the stationary equations:
\begin{subequations}
  \label{stat-cme-1}
  \begin{align}
    \mathrm{i} A_1'(Z) + \Omega A_1 + B_1 + \Gamma ( |A_1|^2 + 2 |B_1|^2 ) A_1 = 0, \\
    - \mathrm{i} B_1'(Z) + \Omega B_1 + A_1 + \Gamma ( 2 |A_1|^2 +
    |B_1|^2 ) B_1 = 0.
  \end{align}
\end{subequations}
For $\Omega \in (-1,1)$, these equations admit the exact solutions:
\begin{subequations}
 \label{gap-soliton}
\begin{align}
  A_1(Z) &=\sqrt{\frac{{2}}{{3 \Gamma}}}
  \frac{\mu}{\alpha \cosh(\mu Z) -\im\beta \sinh(\mu Z)},\\
  B_1(Z) &=\sqrt{\frac{{2}}{{3 \Gamma}}}
  \frac{-\mu}{\alpha \cosh(\mu Z) +\im\beta \sinh(\mu Z)},
\end{align}
\end{subequations}
where
\[
\alpha = \sqrt{1 + \Omega}, \quad \beta = \sqrt{1 - \Omega}, \quad \mu
= \sqrt{1 - \Omega^2} \equiv \alpha \beta.
\]
The localized solution \eqref{gap-soliton} satisfies the symmetry
property
\[
A_1(Z) = \bar{A}_1(-Z), \quad B_1(Z) = \bar{B}_1(-Z), \quad Z \in \R.
\]

We shall call the solution of \eqref{gap-soliton} a {\it monochromatic
  gap soliton}, since the associated approximate solution of the
nonlinear Maxwell model consists of a slowing varying and localized
envelope with a single fast (carrier) frequency of oscillation.  This
is in contrast to the broad band, or polychromatic, solitons which
possess slowly varying envelopes on multiple distinct carrier
frequencies.  It seems unlikely that there is an explicit solution of
the system \eqref{stat-cme-system} of infinitely many coupled mode
equations.

\subsection{Persistence of Solitons in a Band Edge Approximation}

We now explore a small amplitude, spectral band edge, approximation of
xNLCME, which will lead to an infinite system of coupled NLS type
equations, xNLS.  

% Though we are unable to prove the existence of
% stationary solutions in xNLS, this model has a somewhat simplified
% Hamiltonian structure which reveals the loss of compactness making the
% problem challenging.

\subsubsection{The Band Edge Approximation}

The gap in the continuous spectrum exists for $\Omega \in (-1,1)$. In
the truncated coupled mode equations \eqref{stat-cme-1}, the exact
solution \eqref{gap-soliton} shows that the amplitude $\| A_1
\|_{L^{\infty}}$ of the gap soliton becomes small as $\Omega \to 1$.
Using the parameterization $\Omega = \sqrt{1 - \mu^2}$ and the
asymptotic expansion
$$
A_1 = \mu U_1(\zeta) + \bigo(\mu^2), \quad B_1 = -\mu U_1(\zeta) +
\bigo(\mu^2),
$$
where $\zeta = \mu Z$ is slow variable and $\mu$ is a small parameter,
we can formally reduce the system of differential equations
\eqref{stat-cme-1} to the scalar second-order equation for
$U_1(\zeta)$:
\begin{equation}
  \label{normal-form-ode} U_1''(\zeta) - U_1(\zeta) + 6 \Gamma U_1^3(\zeta)
  = 0.
\end{equation}
This equation admits the localized solution
\begin{equation}
  \label{soliton-NLS} U_\star(\zeta) = \frac{1}{\sqrt{3 \Gamma}} {\rm
    sech}(\zeta),
\end{equation}
which corresponds to the asymptotic approximation of the gap soliton
\eqref{gap-soliton} as $\Omega \to 1$.

Generalizing this approach to the system of infinitely many coupled
mode equations, \eqref{stat-cme-system}, we substitute $\Omega =
\sqrt{1 - \mu^2}$ and
\begin{align}
  A_p = \mu {\tilde{A}_p(\zeta) }, \quad B_p = -\mu
  {\tilde{B}_p(\zeta)}, \quad p \in \Zodd,
\end{align}
with $\zeta = \mu Z$ into the coupled mode system
\eqref{stat-cme-system} to obtain
\begin{subequations}
  \label{coupled-mode-tilded}
  \begin{align}
    \mathrm{i} \mu \tilde{A}_p' + p \sqrt{1-\mu^2} \tilde{A}_p - p
    \tilde{B}_p
    + \tfrac{p}{3}\Gamma \mu^2 \tilde{F}_p &=0\\
    \mathrm{i} \mu \tilde{B}_p' - p \sqrt{1-\mu^2} \tilde{B}_p +p
    \tilde{A}_p - \tfrac{p}{3}\Gamma \mu^2 \tilde{G}_p &=0,
  \end{align}
\end{subequations}
where
\begin{subequations}
  \label{e:small_amp_fields}
\begin{align}
  \tilde{F}_p &= 3 \tilde{A}_p \sum_{q \in \Zodd} \abs{\tilde{B}_q}^2 +
  \sum_{q,r\in \Zodd} \tilde{A}_q \tilde{A}_r \tilde{A}_{p-q-r}, \\
  \tilde{G}_p &= 3 \tilde{B}_p \sum_{q \in \Zodd} \abs{\tilde{A}_q}^2 +
  \sum_{q,r\in \Zodd} \tilde{B}_q \tilde{B}_r \tilde{B}_{p-q-r},
\end{align}
\end{subequations}

Introducing the variables
\[
\tilde{U}_p = \frac{\tilde{A}_p + \tilde{B}_p}{2}, \quad \tilde{V}_p =
\frac{\tilde{A}_p - \tilde{B}_p}{2},
\]
the system \eqref{coupled-mode-tilded} can be written as
\begin{subequations}
  \label{coupled-mode-tilded-U-V}
  \begin{align}
    2 p \tilde{V}_p + \mathrm{i} \mu \tilde{U}_p' + \left(
      \sqrt{1-\mu^2} - 1 \right) p \tilde{V}_p +
    \frac{1}{6} \Gamma \mu^2 p (\tilde{F}_p - \tilde{G}_p) &= 0, \\
    \mathrm{i} \tilde{V}_p' + \frac{\sqrt{1-\mu^2} - 1}{\mu} p
    \tilde{U}_p + \frac{1}{6} \Gamma \mu p (\tilde{F}_p + \tilde{G}_p)
    &= 0,
  \end{align}
\end{subequations}
where $\tilde{F}_p$, $\tilde{G}_p$ are rewritten after the
substitution of the new variables.

Now, if we formally expand in powers of $\mu$,
\begin{equation}
  \tilde{U}_p = U_p + \bigo(\mu^1), \quad
  \tilde{V}_p = -\frac{\mathrm{i} \mu}{2 p} U'_p + \bigo(\mu^2),
\end{equation}
we find obtain, at leading order, an infinite system of coupled NLS
type equations, that we deem xNLS:
\begin{equation}
  \label{stat-NLS-system} U_p''(\zeta) - p^2 U_p + \frac{2 p^2}{3}
  \Gamma \left( 3 U_p \sum_{q \in \Zodd} |U_q|^2 + \sum_{q \in \Zodd}
    \sum_{r \in \Zodd} U_q U_r U_{p-q-r} \right) = 0, \quad p \in \Z.
\end{equation}
This can be rewritten as the integro-differential equation
\begin{equation}
  \label{stat-NLS-system-fourier} (\partial_{\zeta}^2 + \partial_{\phi}^2 ) U =
  \frac{2}{3}
  \Gamma \partial_{\phi}^2 \left[ U^3 + 3 \left( \frac{1}{2\pi} \int_{-\pi}^{\pi} |U(\zeta,\theta)|^2 d \theta \right) U \right],
\end{equation}
after introducing the Fourier relations
\[
U(\zeta, \phi) = \sum_{p\in \Zodd} U_p(\zeta) e^{i p \phi}, \quad
U_p(\zeta) = \frac{1}{2\pi} \int_{-\pi}^\pi U(\zeta, \phi) e^{-i p
  \phi} d\phi.
\]
We will now justify the reduction to xNLS, \eqref{stat-NLS-system}.

\subsubsection{Preliminaries}
\label{s:prelim}

We first introduce
appropriate function spaces in which we study the problem.  Let
$\mathbb{T}$ denote the interval $[0,2\pi]$, with endpoints identified
so that functions on $\mathbb{T}$ are understood to be
$2\pi$-periodic. We shall consider functions defined on $\R\times \T$,
admitting the Fourier representation
\begin{equation*}
  U(\zeta,\phi) = \sum U_p(\zeta) e^{\im p \phi}.
\end{equation*}
For any $s$, the function space $X^s$ is defined by
\begin{equation}
  \label{Xsdef}
  X^s \equiv \left\{ U(\zeta,\phi) \in H^s(\R \times \mathbb{T}) :\quad\begin{aligned}
    &\bar{U}(\zeta,\phi) = U(\zeta,\phi),\\
    &\int_{-\pi}^{\pi} U(\zeta,\phi)\cos(2p\phi) d \phi = 0, \;\;
    \forall \zeta \in \R, p\in\N
\end{aligned}\right\}
\end{equation}
and equipped with the norm
\begin{equation}
  \label{space-2}
  \| U \|_{X^s}\equiv \left( \sum_{p \in \mathbb{Z}_{\rm odd}}
    \int_{\R} (p^2 + \xi^2)^s |U_p(\xi)|^2 d\xi \right)^{1/2}.
\end{equation}
We shall frequently go back and forth between the $U$ and
$\{U_p\}_{p\in\Zodd}$ representations of functions in $X^s$.

The Sobolev space $H^s(\R \times \mathbb{T})$ is a Banach algebra with respect to the
pointwise multiplication for any $s > 1$.  Moreover, from the
continuous embeddings $H^s(\R \times \mathbb{T})\hookrightarrow
L^{\infty}(\R \times \mathbb{T})$ for $s > 1$ and
$l^2(\mathbb{Z})\hookrightarrow l^{\infty}(\mathbb{Z})$, we infer that
if $U \in X^s$ for $s > 1$, then
\begin{equation}
  \label{space-decay}
  \lim_{|\zeta| \to \infty} U(\zeta,\phi) = 0, \quad \forall \phi \in \mathbb{T}.
\end{equation}

Let $B_{\delta}(X^s)$ denote a ball of radius $\delta$ in Banach space
$X^s$ centered at the origin. The Hamiltonian $H$ with the density
\eqref{e:constrained_hamiltonian} consists of the terms controlled by the $H^1$
norms of $E^{\pm}$. To see this, recall the continuous embedding
$H^1(\R \times \mathbb{T})\hookrightarrow L^4(\R \times \mathbb{T})$.
It follows that for any $E^{\pm} \in B_{\delta}(X^s)$ with $s \geq 1$,
there is a constant $C_{\delta,s} > 0$ such that
\[
H \leq\ C_{\delta,s} \left( \| E^+ \|_{X^s} + \|
  E^- \|_{X^s} \right).
\]
Furthermore, the map $(E^+,E^-)\mapsto H$ is
continuous in $X^s$. Although we
will mainly study the problem in $X^s$ with $s > 1$, we note that the
energy is well defined in $X^1$.

\subsubsection{Proof of Result}
\label{s:proof}

We now rigorously justify the small amplitude approximation of
\eqref{stat-cme-system} by  \eqref{stat-NLS-system}.

\begin{theorem}
  % \label{thm:smallamp}
  Fix $s > 1$ and assume the existence of localized solution $U \in
  X^s$ to \eqref{stat-NLS-system-fourier} satisfying the reversibility
  symmetry,
  \begin{equation}
    \label{reversibility-NLS} U_p(\zeta) = \bar{U}_p(-\zeta), \quad p
    \in \Zodd, \quad \zeta \in \R.
  \end{equation}
  Also assume that the linearized operator of system
  \eqref{stat-NLS-system-fourier} at $U$ is invertible in the subspace
  of $X^s$ associated with the constraint
  \eqref{reversibility-NLS}.

  There exists $\mu_0 > 0$ such that for any $\mu \in (-\mu_0,\mu_0)$,
  the system of stationary coupled mode equations
  \eqref{stat-cme-system} with $\Omega = \sqrt{1 - \mu^2}$ admits a
  unique localized solution $A, B \in X^s$ satisfying the symmetries,
  \begin{equation}
    \label{reversibility-CMS} A_p(Z) = \bar{A}_p(-Z), \quad B_p(z) =
    \bar{B}_p(-Z), \quad p \in \Zodd, \quad Z \in \R,
  \end{equation}
  and the bound,
  \begin{equation}
    \label{bound-main-2} \| A - \mu U(\mu \cdot,\cdot) \|_{X^s} + \|
    B + \mu U(\mu \cdot,\cdot) \|_{X^s} \leq C \mu^2.
  \end{equation}
  \label{theorem-main-2}
\end{theorem}

\begin{proof}
  First, we note that the vectors fields $\tilde{F}(A,
    B)$ and $\tilde{ G}( A,B )$, defined by their components in
  \eqref{e:small_amp_fields}, are analytic (cubic) maps from $X^s
  \times X^s$ to $X^s$ for any $s > 1$. Eliminating $\tilde{U}_p$ from 
  system \eqref{coupled-mode-tilded-U-V}, we obtain 
  \begin{equation}
  \label{technical-V}
  p^2 \tilde{V}_p - \tilde{V}_p'' =
    \frac{1}{6} \Gamma \bracket{p^2 (\sqrt{1-\mu^2}-1) (\tilde{F}_p - \tilde{G}_p) 
    - i \mu p (\tilde{F}_p' + \tilde{G}_p')}. 
\end{equation}
  The right-hand side of system (\ref{technical-V}) defines an analytic (cubic) map 
  from $X^s$ to $X^{s-2}$ for any $s > 1$, where the $X^{s-2}$ norm is
  of order 
  $\bigo(\mu)$ as $\mu \to \infty$. The left-hand side operator of system (\ref{technical-V}) 
  has a bounded inverse from $X^{s-2}$ to $X^s$, thanks to the zero mean constraint in $X^s$. By the
  Implicit Function Theorem, we infer that for any $\delta > 0$ and any $s > 1$, there is $\mu_0 > 0$ 
  such that for all $\mu \in (-\mu_0,\mu_0)$ and for all
  $\tilde{U}$ in a ball $B_{\delta}(X^s)$, there is a
  smooth map $X^s \ni \tilde{U} \mapsto \tilde{V}[\tilde{U}] \in X^s$
  which solves system \eqref{technical-V} and satisfies
  the bound,
  \begin{equation}
    \label{bound-on-tilde-V} \exists C > 0 : \quad \| \tilde V \|_{X^s} \leq
    C \mu, \quad \mu \in (-\mu_0,\mu_0), \;\; \tilde{U} \in B_{\delta}(X^s).
  \end{equation}

  On the other hand, eliminating $\tilde{V}$ from system
  \eqref{coupled-mode-tilded-U-V}, we obtain 
  \begin{equation}
      \label{system-U-tilde}
    \tilde{U}_p'' - p^2 \tilde{U}_p +
    \frac{1}{6} \Gamma \bracket{p^2 (\sqrt{1-\mu^2} + 1) (\tilde{F}_p + \tilde{G}_p)
    - i \mu p (\tilde{F}_p' - \tilde{G}_p'}=0.
\end{equation}
  Thanks to the bound \eqref{bound-on-tilde-V}, the cubic terms 
  of the system \eqref{system-U-tilde} are different from 
  those of the system \eqref{stat-NLS-system} by the error of the order of ${\cal
    O}(\mu^2)$ in $X^{s-2}$. Under the assumptions of the existence of
  the solution $U \in X^s$ of the truncated coupled NLS equations
  \eqref{stat-NLS-system-fourier} and the invertibility of the
  linearized operator in the subspace of $X^s$ associated with the
  constraint \eqref{reversibility-NLS}, the linearized operator has a
  bounded inverse from $X^{s-2}$ to $X^s$ for any small $\mu \in \R$.
  By the contraction mapping arguments, there is a solution
  $\tilde{U}$ near $U$ in $X^s$ such that
\[
  \exists C > 0 : \quad \| \tilde{U} - U \|_{X^s} \leq C \mu^2.
\]
  This gives the statement of the theorem, after the original
  variables $A$, $B$, and $Z$ are restored from the transformations
  above.
\end{proof}

\subsection{Hamiltonian \& Power of xNLS}

The extended system of coupled nonlinear Schr\"odinger equations
(xNLS) \eqref{stat-NLS-system} inherits 
the Hamiltonian structure of the coupled mode equations
\eqref{stat-cme-system}.  The energy functional for
\eqref{stat-NLS-system} is given by
\begin{equation} \label{hamiltonian-NLS} \small H_{\rm xNLS} =
  \int_{\R} \left[ \sum_{p \in \Zodd} \left( \frac{1}{p^2}
      |U_p'|^2 + |U_p|^2 \right)- \Gamma \left( \sum_{p \in
        \Zodd} |U_p|^2 \right)^2 - \frac{\Gamma}{3} \sum_{p,q,r \in
      \Zodd}
    \bar{U}_p U_q U_r \bar{U}_{q + r - p} \right] d \zeta.
\end{equation}
We also define the power,
\begin{equation} \label{power-NLS} N_{\rm xNLS} = \int_{\R} \left[
    \sum_{p \in \mathbb{Z}} |U_p|^2 \right] d \zeta.
\end{equation}

Energy functionals are often used in proving the existence of
localized solutions to constrained variational problems, {\it e.g.}
\begin{equation}
  \label{var-problem}
  \text{minimize $H_{\rm xNLS}$ subject to fixed $ N_{\rm xNLS}$.}
\end{equation}
Unfortunately, this strategy fails for our problem, as demonstrated by
the following counterexample.  Let
\begin{equation}
  \label{counter-example}
  U_p(\zeta) = \lambda_n^{1/2} W(\lambda_n \zeta) \left( \delta_{p,n} + \delta_{p,-n} \right),
  \quad p \in \Zodd,
\end{equation}
where $W \in H^1(\R)$ is a fixed function, $\lambda_n > 0$ is an
arbitrary parameter, and $n \geq 1$ is an arbitrary odd integer. Then,
$N_{\rm xNLS}$ is independent on the parameters $\lambda_n$ and $n$. On
the other hand,
\[
H_{\rm xNLS} = \frac{2 \lambda_n^2}{n^2} \| W' \|^2_{L^2} - 6 \lambda_n
\| B \|^4_{L^4}.
\]
If we set $\lambda_n = n$ and let $n \to \infty$, we obtain no lower
bound on $H_{\rm xNLS}$.  Thus, localized solutions of xNLS,
\eqref{stat-NLS-system}, if they exist in some $X^s$, cannot be global
minimizers; they will either be local extrema or saddle points.

\subsection{Persistence of Monochromatic Solitons to Coupling in xNLS}

% Another simplification is made by decoupling the components of xNLS,
% \eqref{stat-NLS-system}, through the introduction of an artificial
% parameter. Hence, we formulate xNLS$^\eps$:

We study the question of persistence of NLS solitons within xNLS by
embedding xNLS in a one-parameter family of models, xNLS$^\eps$, for
which xNLS$^0$ is an infinite system of decoupled NLS equations and
xNLS$^1$=xNLS.  Our formulation is:
\begin{equation}
  \label{stat-NLS-system-epsilon} U_p''(\zeta) - p^2 U_p + 6 p^2 \Gamma
  U_p^3 + \frac{2 p^2}{3} \epsilon \Gamma \left( 3 U_p \sum_{q \in
      \Zodd} |U_q|^2 + \sum_{q,r \in \Zodd} U_q U_r U_{p-q-r}
  \right) = 0, \quad p \in \Zodd,
\end{equation}
where the sums exclude the cubic self interaction terms,
$U_p^3$. Within each mode of the decoupled system at $\epsilon = 0$,
we have a solution
\begin{equation}
\label{e:exact_soliton}
U_p(\zeta) = \pm U_\star(p \zeta), \quad p \in \Z_{\rm odd},
\end{equation}
where $U_\star(\zeta)$ is the NLS soliton \eqref{soliton-NLS}.

We now prove the persistence of \eqref{e:exact_soliton} within
xNLS$^{\eps}$, \eqref{stat-NLS-system-epsilon}, for all
$\eps$ sufficiently small.  Without loss of generality,  we can take
$p=1$.  Furthermore, we make the reduction
\[
U_p(\zeta) = \bar U_{p}(\zeta)  = U_{-p}(\zeta), \quad, p \in \Zodd.
\]
In other words, we now assume that the envelopes in each harmonic are
real-valued.

\begin{theorem}
  \label{thm:xnls_mono}
  Fix $s>1$.  There exists $\eps_0 >0$ and $C>0$ such
  that for any $\eps \in (-\eps_0, \eps_0)$, xNLS,
  \eqref{stat-NLS-system-epsilon} admits a unique
  localized solution $U\in X^s$ satisfying the even symmetry:
  \begin{equation}
    \label{e:xnls_symmetry}
    U_p(\zeta) =  U_p(-\zeta)
  \end{equation}

  Moreover, $U(\zeta,\phi)$ is a small deformation of the unperturbed $\eps=0$
  soliton solution:
  \[
  \mathcal{U}_\star(\zeta,\phi)= 2U_\star(\zeta) \cos(\phi),
  \]
  in the sense that
  \begin{equation}
    \label{e:deformation_bound}
    \norm{U - \mathcal{U}}_{X^s} \leq C\eps
  \end{equation}
\end{theorem}

\begin{proof}
  The proof relies on a Lyapunov-Schmidt reduction where we shall
  first express the higher harmonics as functions of $U_{1}=U_{-1}$, and
  then apply the implicit function theorem to an equation written
  entirely in terms of $U_1$.

  From \eqref{stat-NLS-system-epsilon}, define $F$ in terms of the components
  \[
  F_p = 3 U_p \sum_{q \in \Zodd} |U_q|^2 + \sum_{q,r \in \Zodd} U_q U_r
  U_{p-q-r},
  \]
  where each $F_p$ excludes the purely self-interacting terms.  For
  $\abs{p}> 1$, we can clearly write
  \begin{equation}
    \label{e:component_inverse}
  U_p = -(\partial_\zeta^2 - p^2)^{-1}p^2 \paren{6 \Gamma U_p^3} -\eps
  (\partial_\zeta^2 - p^2)^{-1}p^2 \frac{2}{3}\Gamma F_p.
  \end{equation}
  The terms on the left are in $X^s$ since $s>1$  and $(\partial_\zeta^2 - p^2)^{-1}p^2$ is a bounded
  operator.  Therefore, for sufficiently small
  $\eps_0>0$ and finite $\delta_0$, the contraction mapping theorem yields
  a unique map,
  \[
  \Phi: (U_1, \eps) \mapsto \set{U_p}_{\abs{p} >1}
  \]
  in a ball $U_1 \in B_\delta(H^s)$ with $\delta<\delta_0$ and
  $\abs{\eps}<\eps_0$. For a given $U_1 = U_{-1}$, we have expressed
  the other modes in terms of this fixed profile.  Form
  \eqref{e:component_inverse}, we can see that there exists a constant
  $C>0$ such that for all $\abs{\eps}< \eps_0$,
  \[
  \norm{\Phi(U_1, \eps)}_{X^s} \leq C \eps \norm{U_1}_{H^s}^3
  \]

  We now eliminate $\{U\}_{\abs{p}>1}$ from the $p=\pm 1$ equations of
  \eqref{stat-NLS-system-epsilon} using the mapping $\Phi$.  Since $U_1 = U_{-1}$, we
  only consider the $p=1$ equation:
  \begin{equation}
    \label{e:reduced_xnls}
    U_1'' - U_1 + 6 \Gamma U_1^3 = - \frac{2}{3}\eps \Gamma F_1[\Phi(U_1, \eps)].
  \end{equation}
  For any $U_1 \in B_\delta(H^s)$ with finite $\delta>0$ and small $\abs{\eps} < \eps_0$, there is
  a $C>0$ such that
  \[
  \norm{F_1[\Phi(U_1, \eps)]}_{H^s} \leq C \eps \norm{U_1}_{H^s}^5.
  \]

  To solve \eqref{e:reduced_xnls}, we hope to
  apply the implicit function theorem.  Thus, we must linearize
  \eqref{e:reduced_xnls} near $U_\star$ at $\eps=0$, and show that it
  is invertible.  The kernel of the linearized operator,
  \[
  \partial_\zeta^2 - 1 + 18 \Gamma U_\star^2
  \]
  is just $\partial_\zeta U_\star$.  Obviously, this does not satisfy
  the symmetry constraint, \eqref{e:xnls_symmetry}.  Subject to this
  condition, the operator is an isomorphism from $H^s_{\rm even}\to
  H^{s-2}_{\rm even}$, the subspace of $H^s$ of even functions. Hence, the implicit function
  theorem yields a neighborhood of $U_\star$ in
  $H^s$ in which we can obtain $U_1$ with $\abs{\eps}< \eps_1 \leq
  \eps_0$.

  Moreover, we see from \eqref{e:reduced_xnls} that there exists $C>0$
  such that for all  $\abs{\eps}< \eps_1$,
\[
\norm{U_1 - U_\star}_{H^s} \leq C \eps^2
\]
Combining this estimate with \eqref{e:component_inverse}, yields
\eqref{e:deformation_bound}.

\end{proof}

This result has several obvious extensions.  We can consider the local
continuation about a soliton localized in any other mode $p_0 \in\Zodd$,
\begin{equation}
  U_{p_0}(\zeta) = \pm U_\star(p\zeta).
\end{equation}
We could also continue a solution about any finite collection of such
solitons.  However, if we begin with solitons in every odd harmonic,
they will not have finite $L^2$, as
\[
\int \sum_{p\in \Zodd} \abs{U_p}^2 d\zeta = \norm{U_\star}_{L^2}^2
\sum_{p\in \Zodd} \frac{1}{\abs{p}}
\]
which diverges.  The continuation of such infinite energy solutions is
an open problem.

\section{Variational Approximations}
\label{s:var_approx}

As noted at the end of Section \ref{s:proof}, although the functional
is bounded from below, the natural variational formulation for localized solutions of xNLS
exhibits a loss of compactness.  In this section we explore the use of
this functional to obtain Rayleigh-Ritz or Galerkin-type
approximations to such states. The parameters in these approximations
can be uniquely determined from the conditions that the ansatz gives a
stationary point of $H_{\rm xNLS}$. Indeed, variations of $H_{\rm
  xNLS}$ produce the Euler--Lagrange equations, which are equivalent
to the differential equations \eqref{stat-NLS-system}.

\subsection{Gaussian Approximations}

Let us consider the Gaussian variational ansatz
\begin{equation}
  \label{Gaussian-ansatz}
  U_p(\zeta) = a_p e^{-b_p \zeta^2}, \quad p \in \Z_{\rm odd},
\end{equation}
where $a_p \in \R$ and $b_p \in \R_+$ are parameters of the
variational approximation.  The Gaussian ansatz is useful because all
integrals in $H_{\rm xNLS}$ can be computed in the analytical form.
Direct substitution and integration show that
$\sqrt{\tfrac{2}{\pi}}H_{\rm xNLS}$ becomes
\begin{equation} \label{hamiltonian-ansatz} H_{\rm Gauss} = \sum_{p
    \in \mathbb{Z}_{\rm odd}} \frac{\sqrt{b_p} a_p^2}{p^2} +
  \frac{a_p^2}{\sqrt{b_p}} - \Gamma \sum_{p,q \in \mathbb{Z}_{\rm
      odd}} \frac{a_p^2 a_q^2}{\sqrt{b_p+b_q}} - \frac{1}{3} \Gamma
  \sum_{p,q,r \in \mathbb{Z}_{\rm odd}}\frac{\sqrt{2} a_p a_q a_r
    a_{p-q-r}}{\sqrt{b_p + b_q + b_r + b_{p-q-r}}}.
\end{equation}

If the artificial small parameter $\epsilon$ is introduced to decouple
the different modes, as in
system \eqref{stat-NLS-system-epsilon}, then
\eqref{hamiltonian-ansatz} is rewritten in the form,
\begin{equation}
  \begin{split}
    H_{\rm Gauss}(\eps) &\equiv \sum_{p \in \Zo} \frac{\sqrt{b_p}
      a_p^2}{p^2} + \frac{a_p^2}{\sqrt{b_p}} -
    \Gamma\frac{3 a_p^4}{\sqrt{2}\sqrt{b_p}}\\
    &\quad- \eps \Gamma \left( \sum_{p, q\in \Zo} \frac{a_p^2
        a_q^2}{\sqrt{b_p+b_q}} + \frac{1}{3} \sum_{p,q,r \in \Zo}
      \frac{\sqrt{2} a_p a_q a_r a_{p-q-r}}{\sqrt{b_p + b_q + b_r +
          b_{p-q-r}}} \right).
  \end{split} \label{hamiltonian-ansatz-more}
\end{equation}
The above sums with $\eps$ as a prefactor exclude the purely self
interacting $a_p^4/\sqrt{b_p}$ terms.

If $\eps = 0$, there exists an uncoupled solution of the
Euler--Lagrange equations produced from variations of $H_{\rm
  Gauss}(0)$,
\begin{equation}
  \label{var-solution}
  a_p = \pm \frac{2^{3/4}}{3 \Gamma^{1/2}}, \quad b_p = \frac{p^2}{3}, \quad p \in \Z_{\rm odd}.
\end{equation}
The exact solution \eqref{var-solution} will be used as a seed point
in the numerical continuation algorithm.

\subsection{Numerical Continuation}
\label{s:gaussian_continuation}

Truncating $H_{\rm Gauss}(\eps)$ in \eqref{hamiltonian-ansatz-more} to
resolve only $N$ harmonics, we define $H^N_{\rm Gauss}(\mathbf{a},
\mathbf{b},\eps)$.  The associated system of $2N$ Euler-Lagrange
equations is
\begin{equation}
  \label{e:rr_euler_lagrange}
  \nabla_{\bf a} H^N_{\rm Gauss}(\mathbf{a}, \mathbf{b},\eps)  =
  0,\quad \nabla_{\bf b} H^N_{\rm Gauss}(\mathbf{a}, \mathbf{b},\eps)  = 0.
\end{equation}

We now seek solutions of the $\eps=0$ system, where all modes are
decoupled, that can be continued to $\eps=1$, the desired system.  The
natural family of solutions is given by \eqref{var-solution}.  Thus,
for our $\eps=0$ starting point, we consider solutions of the form
\begin{align}
  \label{stationary-a} {\bf a}_* & = \frac{2^{3/4}}{3
    \Gamma^{1/2}} \left( \sigma_1, \sigma_2, \ldots, \sigma_N\right)\\
  \label{stationary-b} {\bf b}_* & = \frac{1}{3} \left( 1^2, 3^2,
    \ldots, (2N-1)^2 \right)
\end{align}
where $\sigma_j \in \left \{-1, 0, 1\right\}$. The variances, ${\bf
  b}_*$, are unaffected by $\boldsymbol{\sigma}$.  Indeed, for
$\sigma_j =0$, $b_j$ is ill-defined, and can take any value.

We now explore continuations from various
$\boldsymbol{\sigma}$'s. Before giving the results, we state the
conjecture that our computations suggest:
\begin{conj}
  For any $N \geq 1$, there is a nontrivial configuration $\boldsymbol{\sigma}$ that can be continued from $\eps =0$
  to $\eps = 1$.  At $\eps=1$, the amplitudes are sign alternating,
 $$
 \sign(a_p) =(-1)^{(\abs{p}-1)/2}.
 $$
\end{conj}

\begin{figure}
  \centering
  \subfigure{\includegraphics[width=3in]{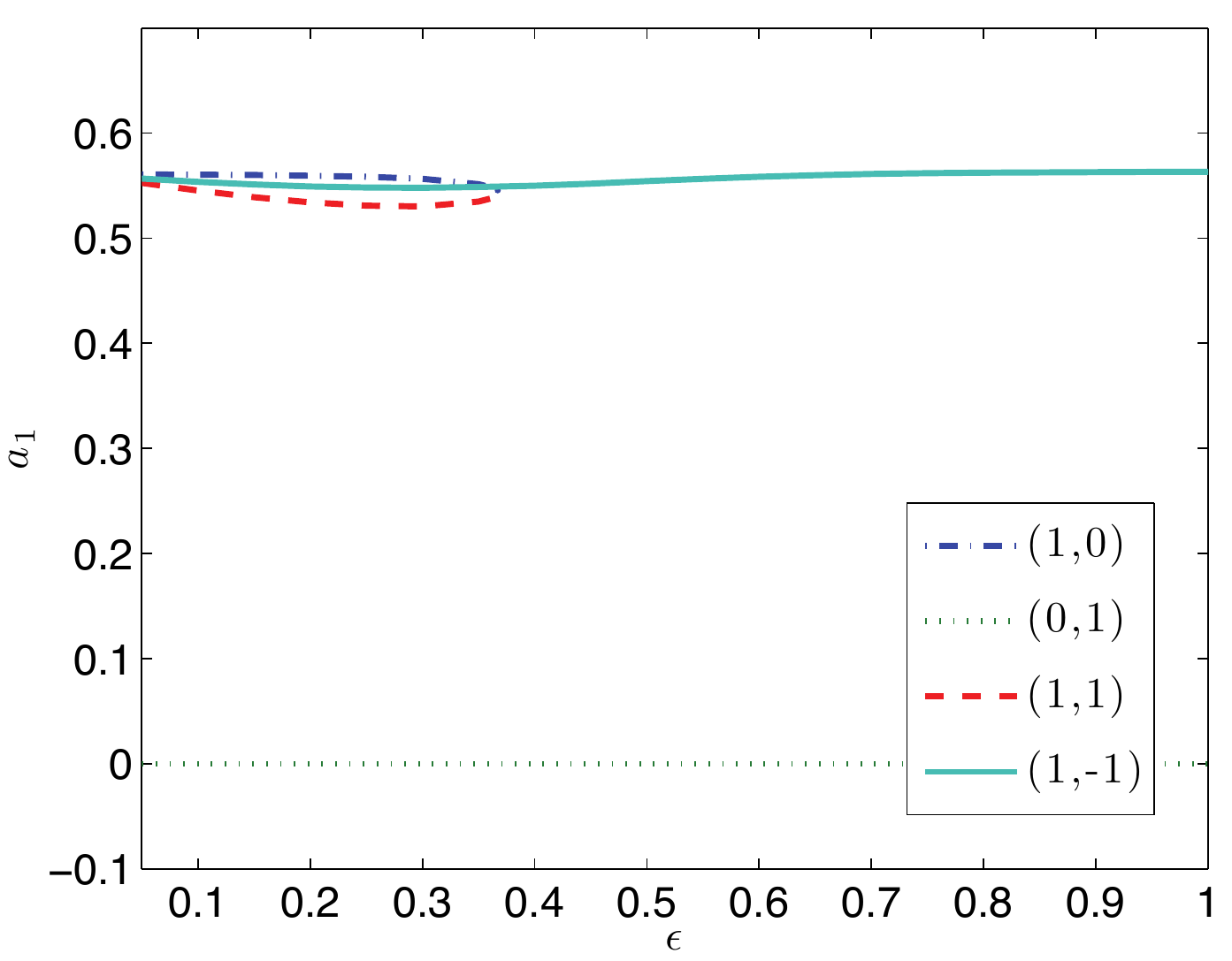}}
  \subfigure{\includegraphics[width=3in]{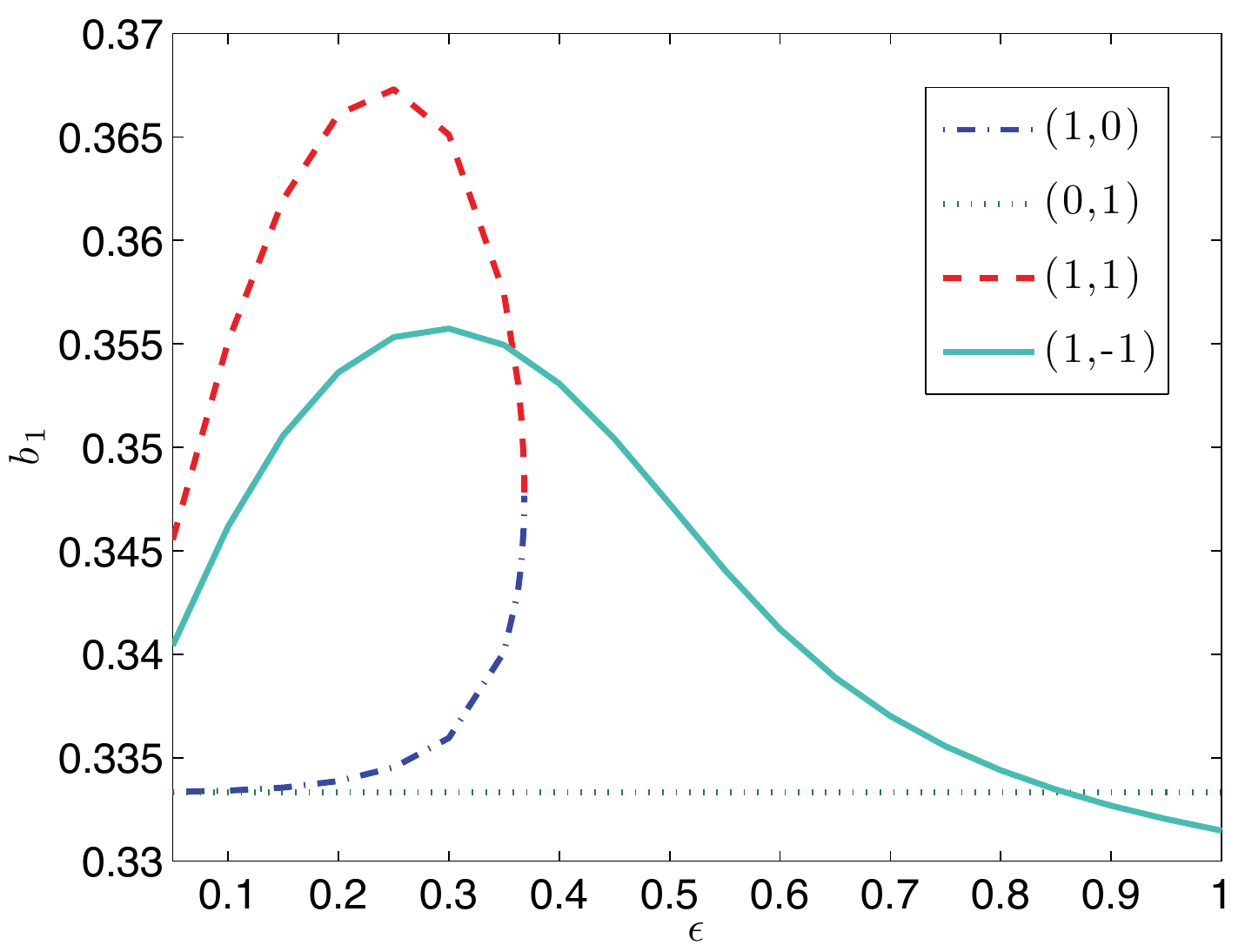}}

  \subfigure{\includegraphics[width=3in]{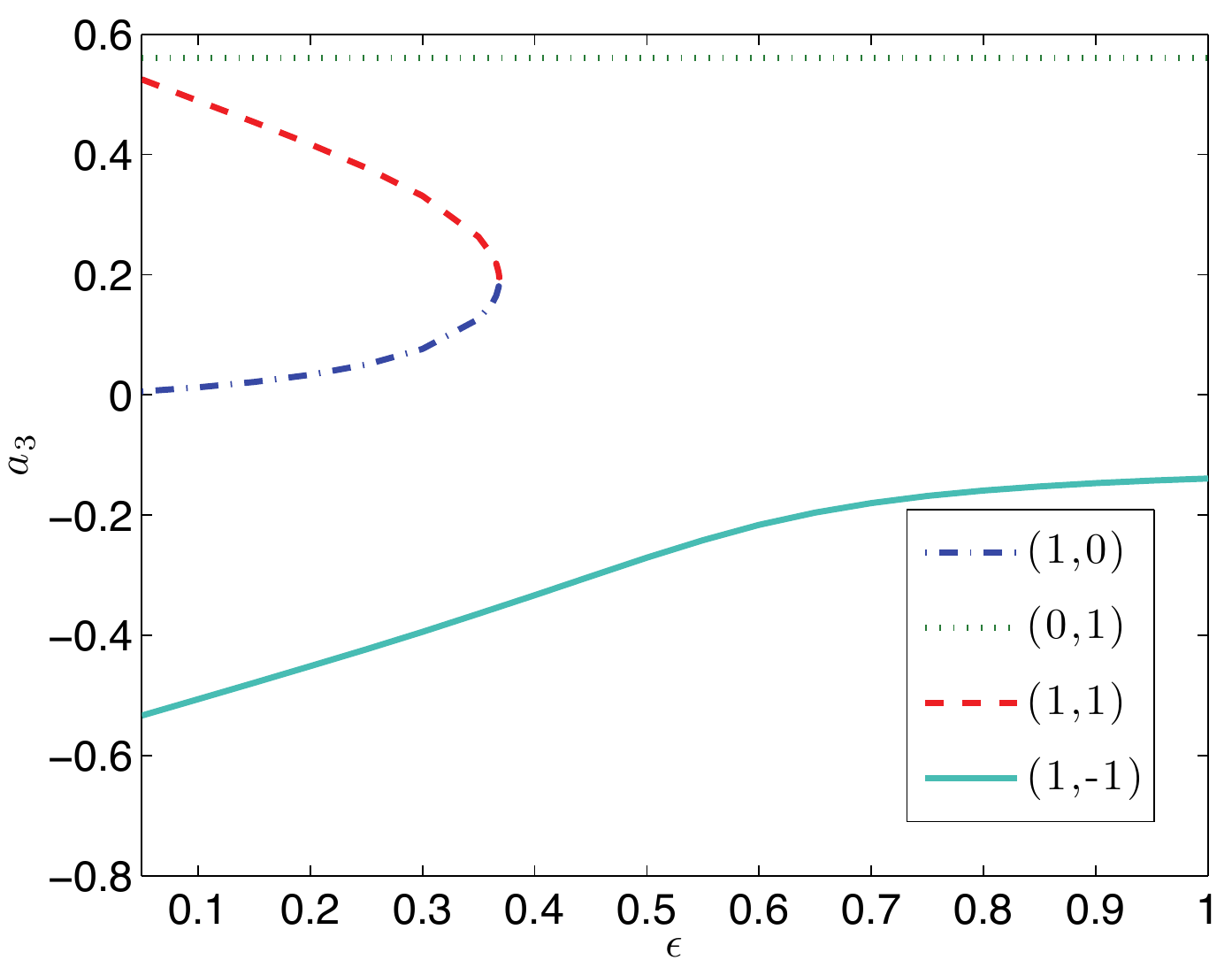}}
  \subfigure{\includegraphics[width=3in]{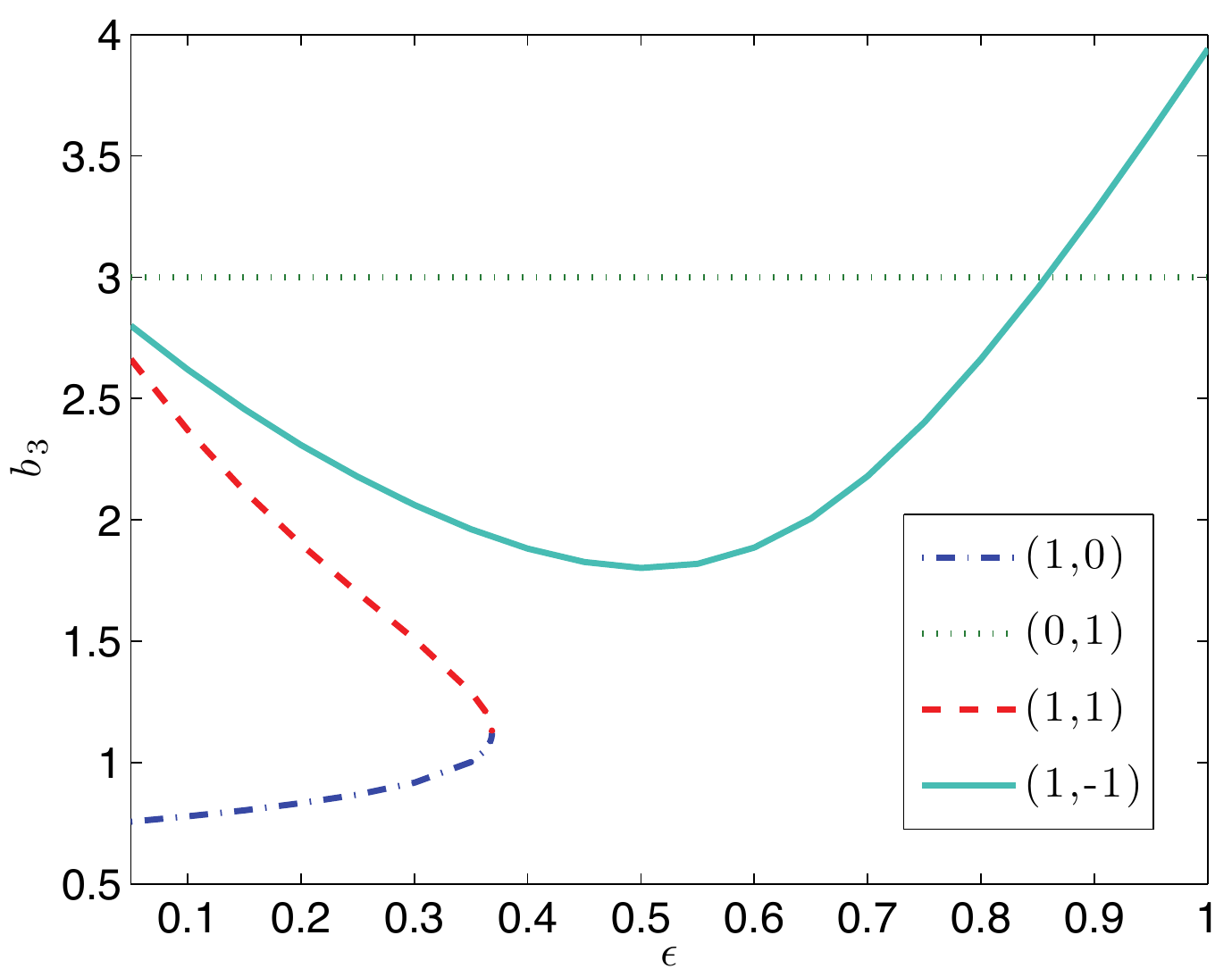}}
  \caption{Various continuation branches for a two-mode Rayleigh-Ritz
    system.}
  \label{f:twomode_rr}
\end{figure}

\begin{figure}
  \centering
  \subfigure{\includegraphics[width=3in]{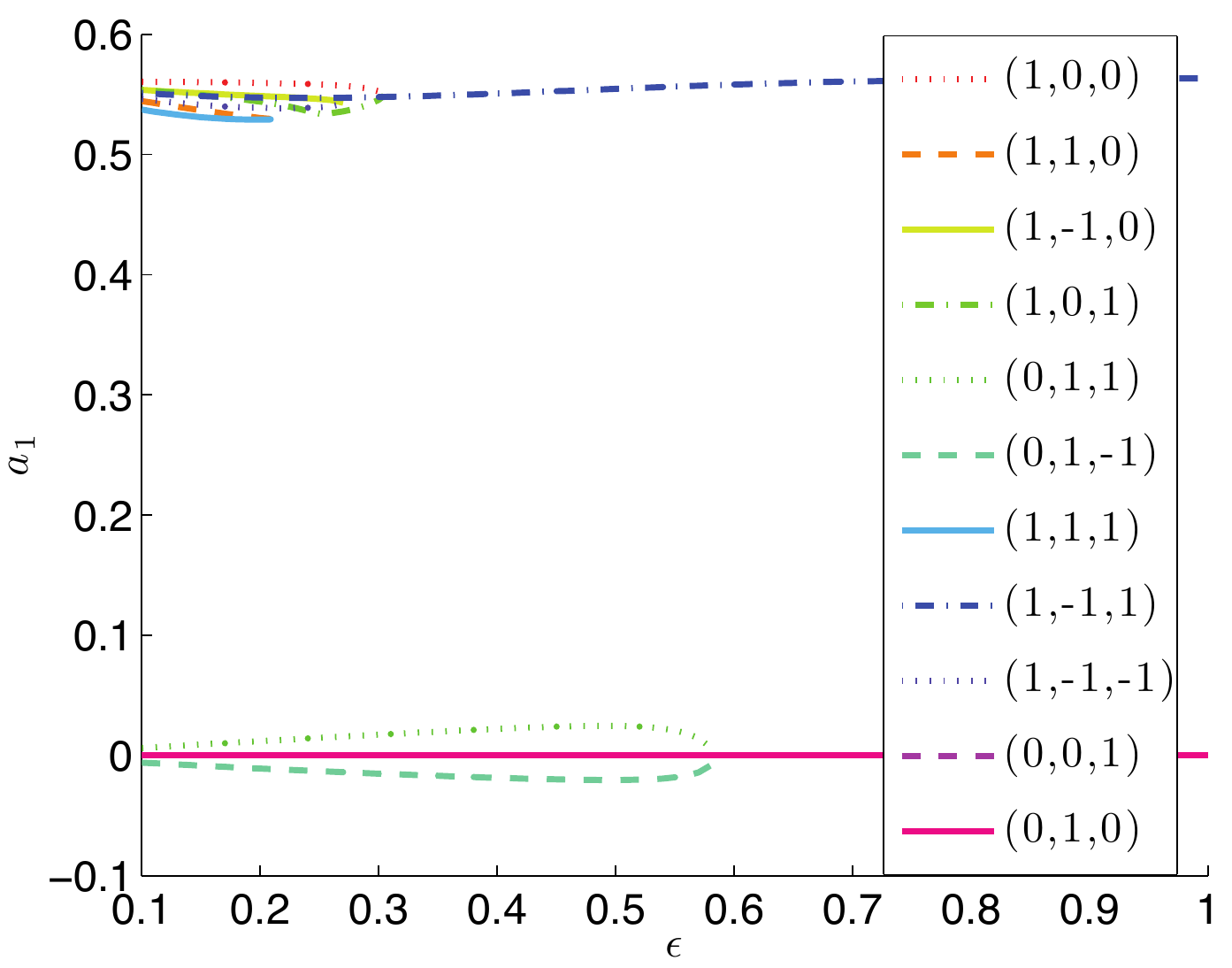}}
  \subfigure{\includegraphics[width=3in]{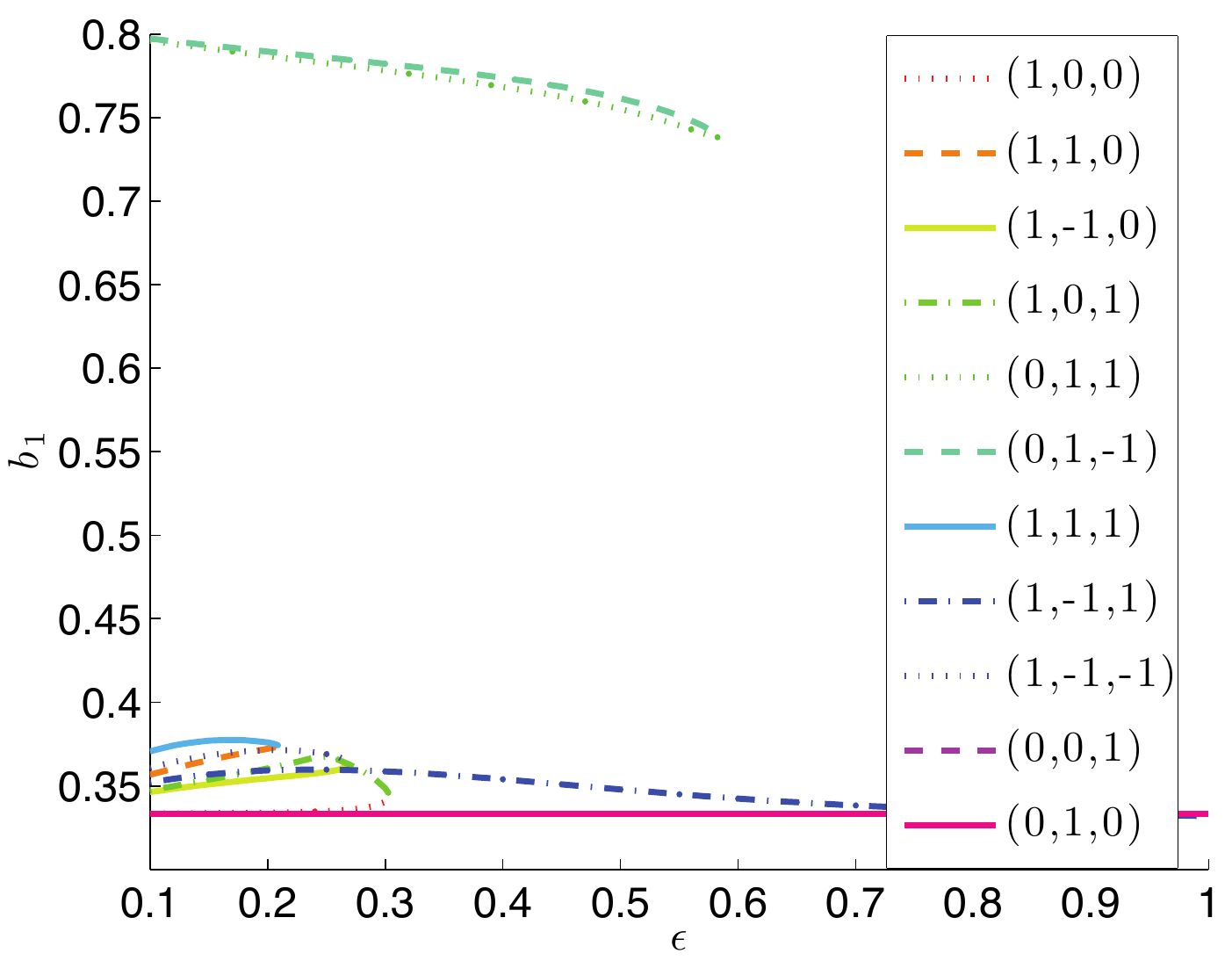}}

  \subfigure{\includegraphics[width=3in]{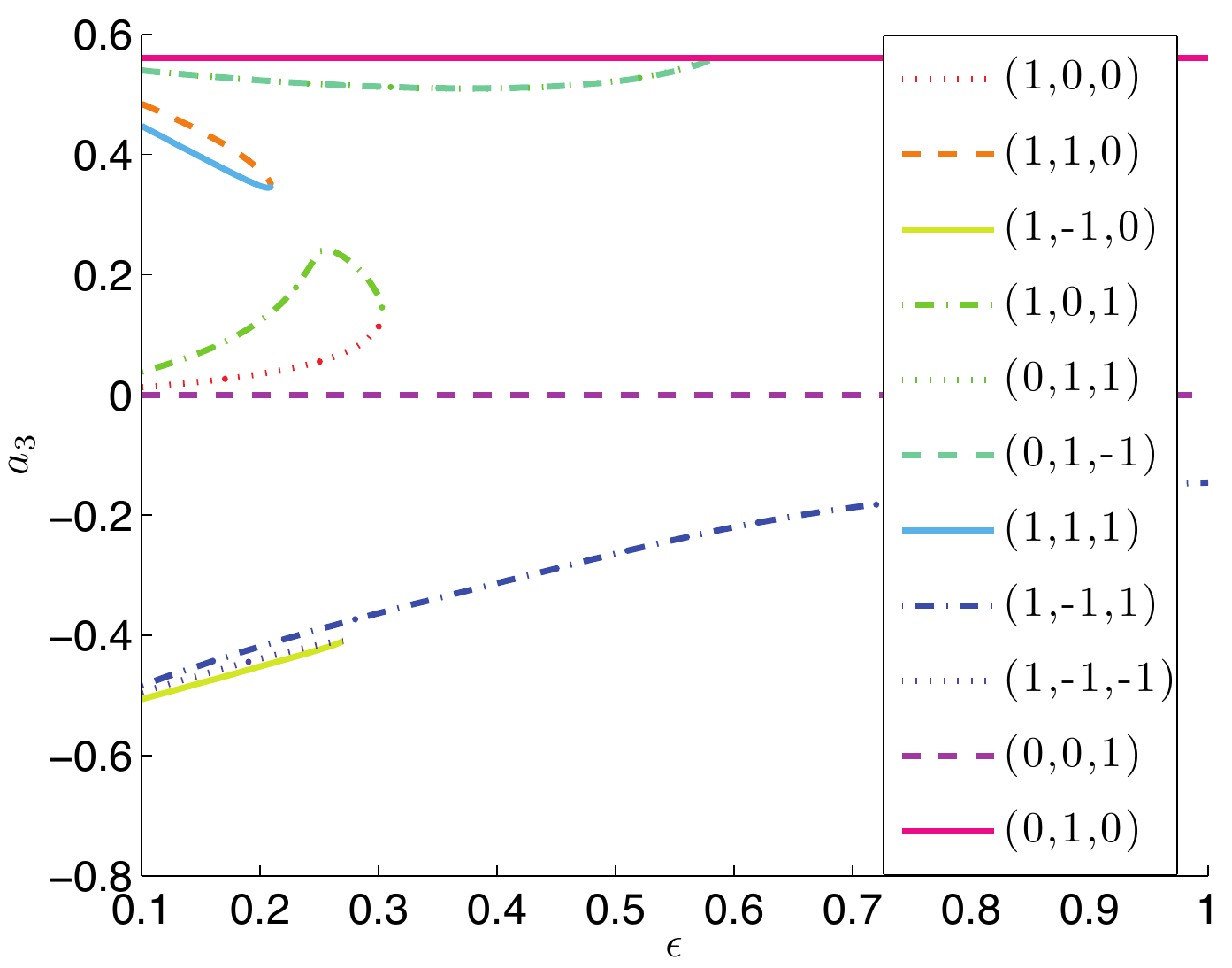}}
  \subfigure{\includegraphics[width=3in]{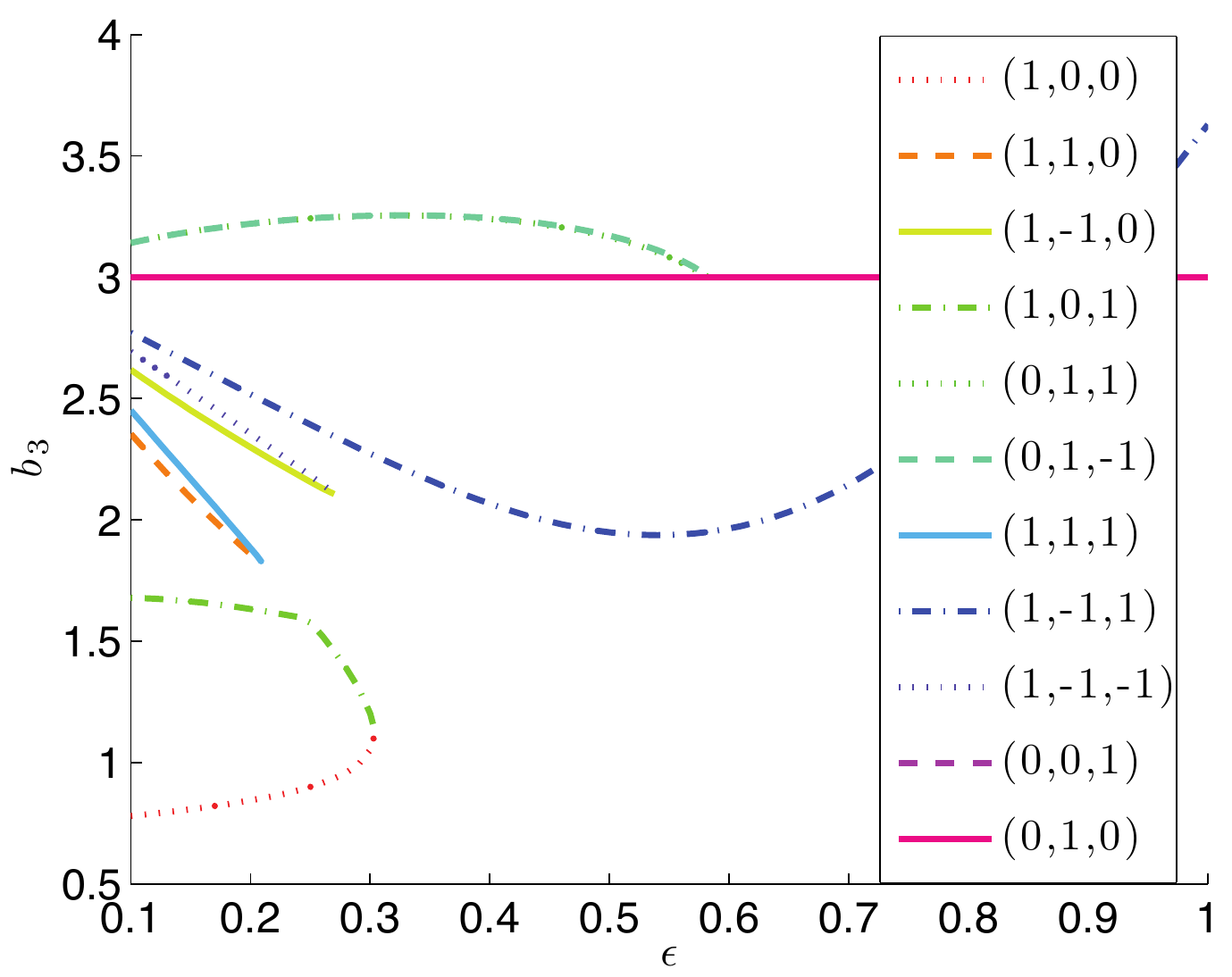}}

  \subfigure{\includegraphics[width=3in]{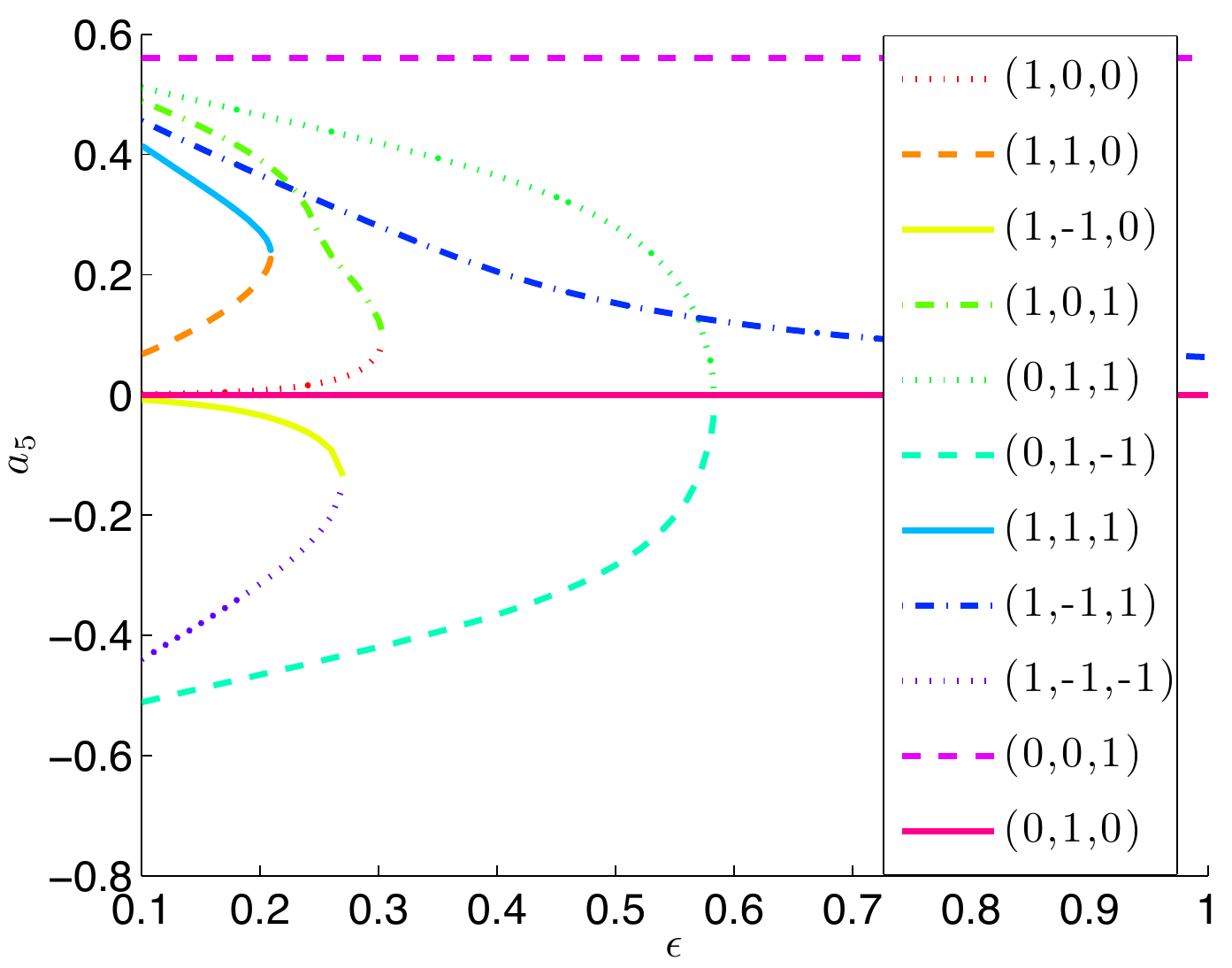}}
  \subfigure{\includegraphics[width=3in]{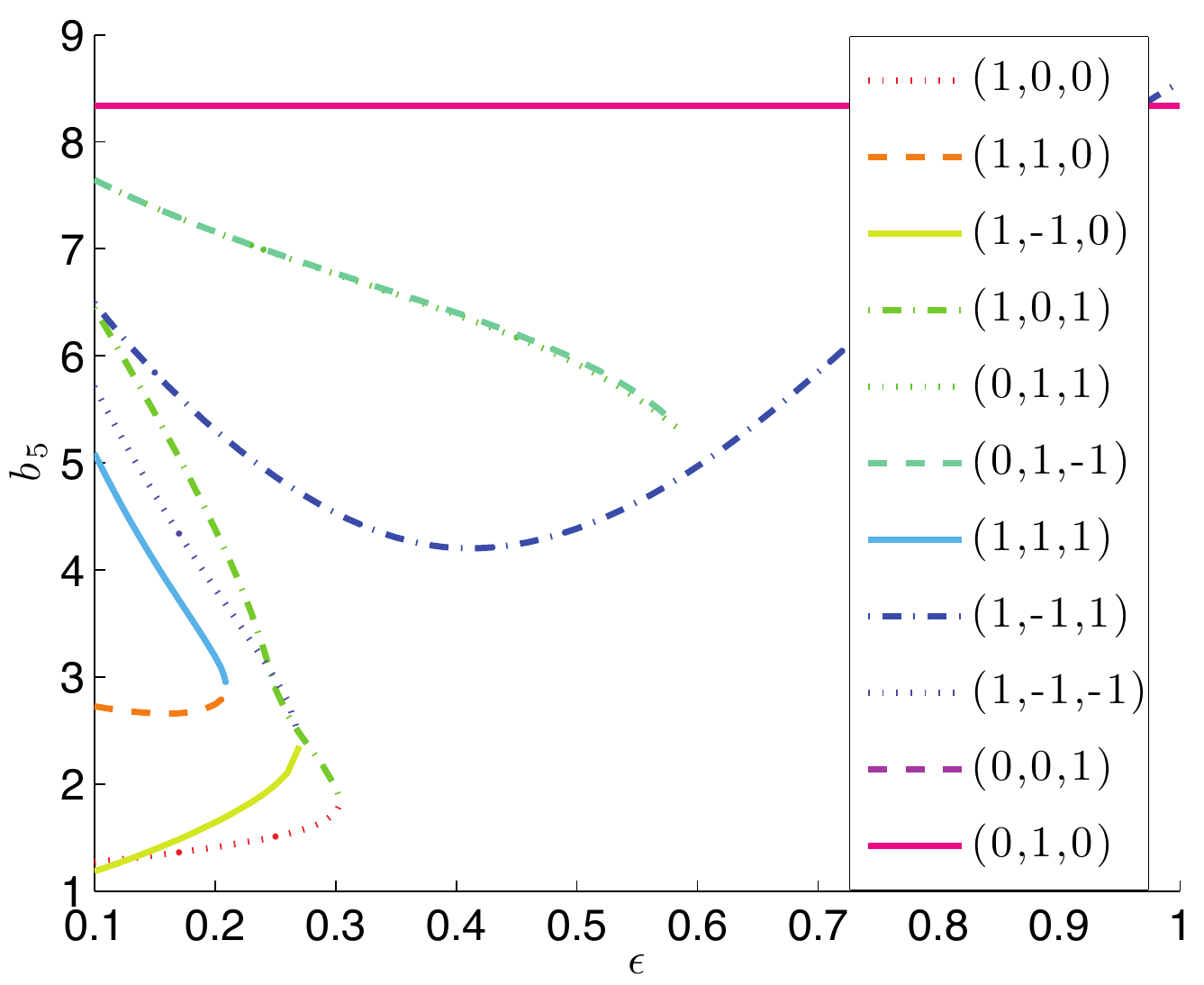}}

  \caption{Various continuation branches for a three-mode
    Rayleigh-Ritz system.}
  \label{f:threemode_rr}
\end{figure}

For a system of two modes ($N=2$), the numerical continuation of four
$\boldsymbol{\sigma}$ configurations is plotted in Figure
\ref{f:twomode_rr}.  The configurations $\boldsymbol{\sigma}=(0,1)$
and $\boldsymbol{\sigma}=(1,-1)$, can be continued to $\eps=1$, while
the other two collide and terminate near $\eps =0.368$.  Extending
this to a system of three modes, we plot the analogous results in
Figure \ref{f:threemode_rr}. Three configurations
$\boldsymbol{\sigma}=(0,1,0)$, $\boldsymbol{\sigma}=(0,0,1)$, and
$\boldsymbol{\sigma}=(1,-1,1)$ can be continued to $\eps=1$.  We note
that the configurations $\boldsymbol{\sigma}=(0,1)$,
$\boldsymbol{\sigma}=(0,1,0)$, and $\boldsymbol{\sigma}=(0,0,1)$ are
trivial in the sense that they are generated by the reductions of the
truncated coupled NLS equations. When more modes are included into the
system, these degenerate configurations are destroyed. On the other
hand, the configurations $\boldsymbol{\sigma} = (1,-1), (1,-1,1)$
are non-trivial and persist with respect to adding more modes in the
coupled NLS equations. Our results for the non-trivial configurations
at $\eps = 1$ are summarized in Table \ref{t:rr_full}.

\begin{table}
  \centering
  \caption{Computed values for a truncated Rayleigh-Ritz
    approximation for $\eps = 1$.}
  \label{t:rr_full}
  \begin{tabular}{@{}l cc cc cc cc cc@{}} \hline
    No. of Modes & $a_1$ & $b_1$ & $a_3$ & $b_3$ & $a_5$ & $b_5$\\
    \hline
    1 & 0.56060 & 0.33333 &- & -&- &-\\
    2 & 0.56321&0.33148 & -0.13918& 3.9413& -&-\\
    3 & 0.56329& 0.33189 & -0.14585& 3.6287&0.062822 &8.5577\\
    \hline
  \end{tabular}
\end{table}

Though computations for two and three modes suggest that an
alternating configuration of $\pm 1$'s can always be successfully
continued to $\eps =1$, this is not the case, as the following
computations demonstrate.  We first make the additional
simplification, observing that the values of $b_j$ in Table
\ref{t:rr_full} are close to $j^2/3$.  This motivates fixing them as
such, and only solving the problem for the amplitudes, ${\bf a}$.
Thus, we solve
\begin{equation}
  \label{e:rr_euler_lagrange_reduced}
  \nabla_{\bf a} H_{\rm Gauss}^N({\bf a}, {\bf b}_\ast, \eps)=0
\end{equation}
where ${\bf b}_\ast$ is given by \eqref{stationary-b}.  The results of
our computations with these fixed variances are given in Table
\ref{t:rr_bfixed}.  Continuation from the alternating branch
$\boldsymbol{\sigma}= (1,-1), (1,-1,1),\ldots$ is successful till
$N=4$. The alternating branch cannot be continued to $\eps=1$ for five
and six modes, though there are other initial states that can be
continued to $\eps=1$, with sign sign alternations at $\eps=1$; see
Table \ref{t:rr_bfixed}.

Though these results were initially computed using a
naive continuation algorithm in {\sc Matlab}, solving with a given
value of $\eps$ and using that solution as the initial guess for a
larger value of $\eps$, they were confirmed by our computations using
AUTO \cite{doedel81, auto07}.

Though the starting branch may not have an alternating sign structure,
sign alternating solutions may still be found at $\eps=1$.  This makes it
challenging to perform numerical continuation with these branches if
we no longer assume the variances to be fixed.  For a system of five
modes, $a_7$ must change sign. When it crosses zero, the variance
becomes ill-defined introducing numerical difficulties. { On the other
  hand, if we iterate the system \eqref{e:rr_euler_lagrange} near the
  solution of \eqref{e:rr_euler_lagrange_reduced} for $\eps = 1$, the
  convergence is usually achieved with few iterations.}

\begin{table}
  \centering
  \caption{Computed values for a truncated Rayleigh-Ritz
    approximation with fixed ${\bf b}_*$ for $\eps = 1$.  The branch from which
    we continue is alternating for $1 \leq N \leq 4$: $\boldsymbol{\sigma}=(1),
    (1,-1),(1,-1,1), (1,-1,1-1)$.  The case $N=5$ is continued from the
    branch $(1, -1, 1, 1, 1)$ and $N=6$ is continued from $(1, -1, 1, -1, -1,-1)$.}
  \label{t:rr_bfixed}
  \begin{tabular}{@{}l c c c c c c @{}} \hline
    No. of Modes & $a_1$ & $a_3$ & $a_5$ & $a_7$ &$ a_9$ & $a_{11}$ \\
    \hline
    1 & 0.56060 & -  &- & - & - & - \\
    2 & 0.5643&  -0.12734& - & -& - & - \\
    3 & 0.56409& -0.13759  &0.061454 & -& - & -\\
    4 & 0.56386 & -0.14037 & 0.068618 & -0.037695 & - & -\\
    5 & 0.56372 & -0.14139& 0.071254& -0.042822& 0.026041 & - \\
    6 &0.56364 &-0.14184 &0.072457 & -0.045015&0.029896 & -0.019323\\
    \hline
  \end{tabular}
\end{table}

\subsection{Tails of the Variational Solutions}
\label{s:tails}

Though we are able to construct a sequence of Rayleigh--Ritz
approximations with Gaussian ansatz, it is not yet clear if such
solutions should exist in space $X^s$ for $s > 1$ or at least have
finite power ($L^2$) in the limit $N \to \infty$.  Indeed, the
solution $({\bf a}_*,{\bf b}_*)$ given by \eqref{var-solution} for
$\epsilon = 0$ with all $a_p \neq 0$ has infinite power, since
\[
\sqrt{\tfrac{2}{\pi}}\int_{\R} \abs{a_p \exp(-b_p \zeta^2)}^2 d\zeta =
\paren{\frac{2}{3}}^{3/2}\frac{1}{\abs{p}\Gamma^2}
\]
and $\sum_{p \in \Z_{\rm odd}} \frac{1}{\abs{p}} = \infty$.  However, the
results of Table \ref{t:rr_bfixed} show that at $\eps=1$, the
sign-alternating amplitudes $\{ a_p \}_{p \in \Zo}$ are also decaying
in $p \in \Z_{\rm odd}$.  We explore whether or not the decay is
sufficiently rapid to have finite power and to belong to the energy
space, where $H_{\rm Gauss}$ is finite. To this end, we employ a more
refined trial-function ansatz, allowing for weak decay of $a_p$:
\begin{equation}
  \label{e:two_parameter_ansatz}
  a_p = A(-1)^{(\abs{p}-1)/2} \abs{p}^{-\gamma} , \quad b_p =
  \tfrac{p^2}{3}, \quad p \in \Zo
\end{equation}
where $A$ and $\gamma$ are unknown parameters to be determined from
the Euler-Lagrange equations. If $\gamma>0$, the Rayleigh-Ritz
approximation has both $H_{\rm Gauss}$ and $N_{\rm Gauss}$ finite.

Substituting \eqref{e:two_parameter_ansatz} into
\eqref{hamiltonian-ansatz} yields a two parameter Hamiltonian
\begin{equation}
  \label{approximation-h}
  h(\gamma,A) =A^2 f(\gamma) - A^4 \Gamma g(\gamma),
\end{equation}
where
\begin{subequations}
  \begin{align}
    \nonumber
    f(\gamma) & = \sum_{p \in \Z_{\rm odd}} \frac{4}{\sqrt{3}} \abs{p}^{-1-2\gamma} \\
    \nonumber
    \begin{split}
      g(\gamma) & =\sum_{p,q \in \Z_{\rm odd}}
      \sqrt{3}\frac{p^{-2\gamma}q^{-2\gamma}}{\sqrt{p^2 +q^2}} \\
      &\quad+ \sum_{p,q,r \in \Z_{\rm
          odd}}\sqrt{\frac{2}{3}}(-1)^{(\abs{p} + \abs{q} + \abs{r} +
        \abs{p-q-r})/2}
      \frac{\abs{p}^{-\gamma}\abs{q}^{-\gamma}\abs{r}^{-\gamma}\abs{p-q-r}^{-\gamma}}{\sqrt{p^2
          +q^2+r^2 + (p-q-r)^2}}
    \end{split}
  \end{align}
\end{subequations}
Solving $\partial_A h(\gamma,A) = 0$, we find
\begin{equation*}
  A^2(\gamma) = \frac{f(\gamma)}{2 \Gamma g(\gamma)}
\end{equation*}
Plugging back in, we get
\begin{equation}
  \label{e:two_parameter_reduced}
  \tilde{h}(\gamma) = h(\gamma, A(\gamma)) = \frac{f(\gamma)^2}{2\Gamma
    g(\gamma)} - \frac{f(\gamma)^2}{4 \Gamma g(\gamma)} =
  \frac{1}{4\Gamma}\frac{f(\gamma)^2}{g(\gamma)}
\end{equation}

\begin{figure}
  \centering
  \includegraphics[width=3in]{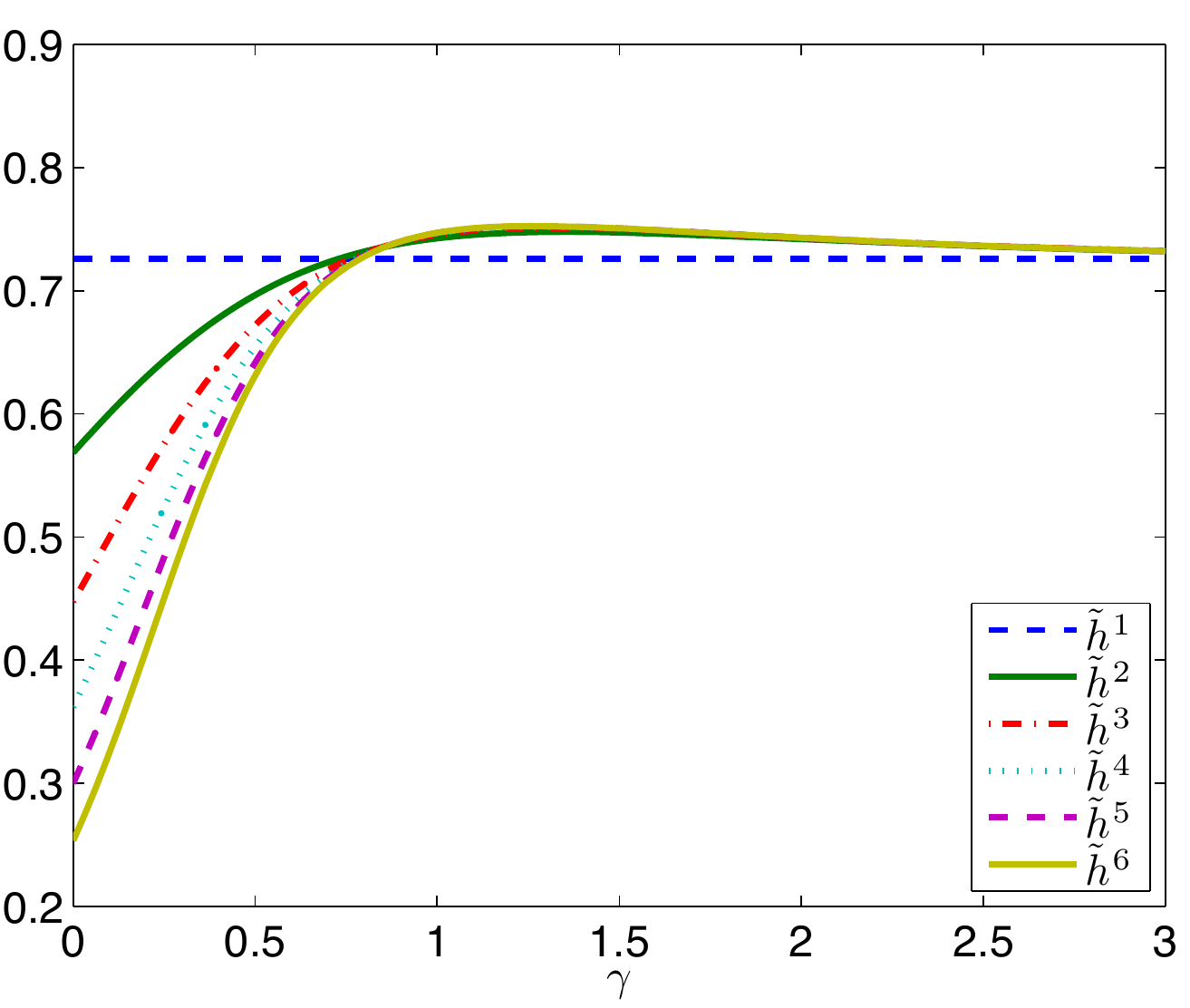}
  \caption{Two--parameter approximation \eqref{e:two_parameter_ansatz}
    of $H_{\rm Gauss}$ for different truncations. }
  \label{f:two_parameter}
\end{figure}

Truncating this approximation to $N$ modes, $\tilde{h}^N(\gamma)$, we
are able to identify a sequence of critical points, suggesting
convergence as $N \to \infty$ and the existence of a critical point in
the primitive functional, \eqref{e:two_parameter_reduced}.  A few of
these approximations are plotted in Figure \ref{f:two_parameter} with
$\Gamma = 1$.
% The first two are:
% \begin{align}
%   \tilde{h}^1(\gamma) &= \frac{16}{9}\sqrt{\frac{2}{3}}\\
%   \tilde{h}^2(\gamma) &=\frac{80 \sqrt{2} 3^{-4 \gamma -2}
%   \left(1+3^{2 \gamma +1}\right)^2}{45 \sqrt{3}+5 3^{\frac{3}{2}-4
%   \gamma }-20 3^{-\gamma }+8 3^{\frac{3}{2}-2 \gamma } \sqrt{5}}
% \end{align}
All of the computed $\tilde{h}^N(\gamma)$'s have the property that
\begin{equation}
  \lim_{\gamma \to \infty}\tilde{h}^N(\gamma) =  \tilde{h}^1(\gamma) = \frac{8}{9}\sqrt{\frac{2}{3}}
\end{equation}
The critical values of $\gamma$, $\gamma_\star$, are given in Table
\ref{t:two_parameter}.  These appear to converge to a value of
$\gamma$ near $\gamma = 1.26$ indicating that the corresponding
variational approximations belong to the energy space of the coupled
NLS equations. Moreover, since
\[
\| U \|_{X^s} \sim \sum_{p \in \Zo} |p|^{2s-1} |a_p|^2 \sim \sum_{p
  \in \Zo} |p|^{2s-1-2\gamma}
\]
and $\gamma \approx 1.26$, the corresponding variational
approximations belong to the space $X^s$ for $s < \gamma$.  Therefore,
the results of Theorems \ref{theorem-main-2} and \ref{thm:xnls_mono}
can be used in the nonempty interval for the values of $s \in
(1,\gamma)$.  As it appears that $\gamma$ is strictly greater than
one as the number of modes increases, a solution of infinitely many modes
might be more regular than $H^1$; indeed, it would be H\"older
continuous.  

\begin{table}
  \centering
  \caption{Computed critical values of $\gamma$ for the curves in
    Figure \ref{f:two_parameter}.}
  \label{t:two_parameter}
  \begin{tabular}{@{}l l l l @{}} \hline
    No. of Modes  &$\gamma_\star$ & $\Delta \gamma_\star$ \\
    \hline
    2 &1.35511&- \\
    3 & 1.30184& 0.05327\\
    4 &  1.28176&0.02008  \\
    5 & 1.27208&0.00968 \\
    6 &1.26672 &0.00536 \\
    \hline
  \end{tabular}
\end{table}

\begin{figure}
  \centering
  \includegraphics[width=3in]{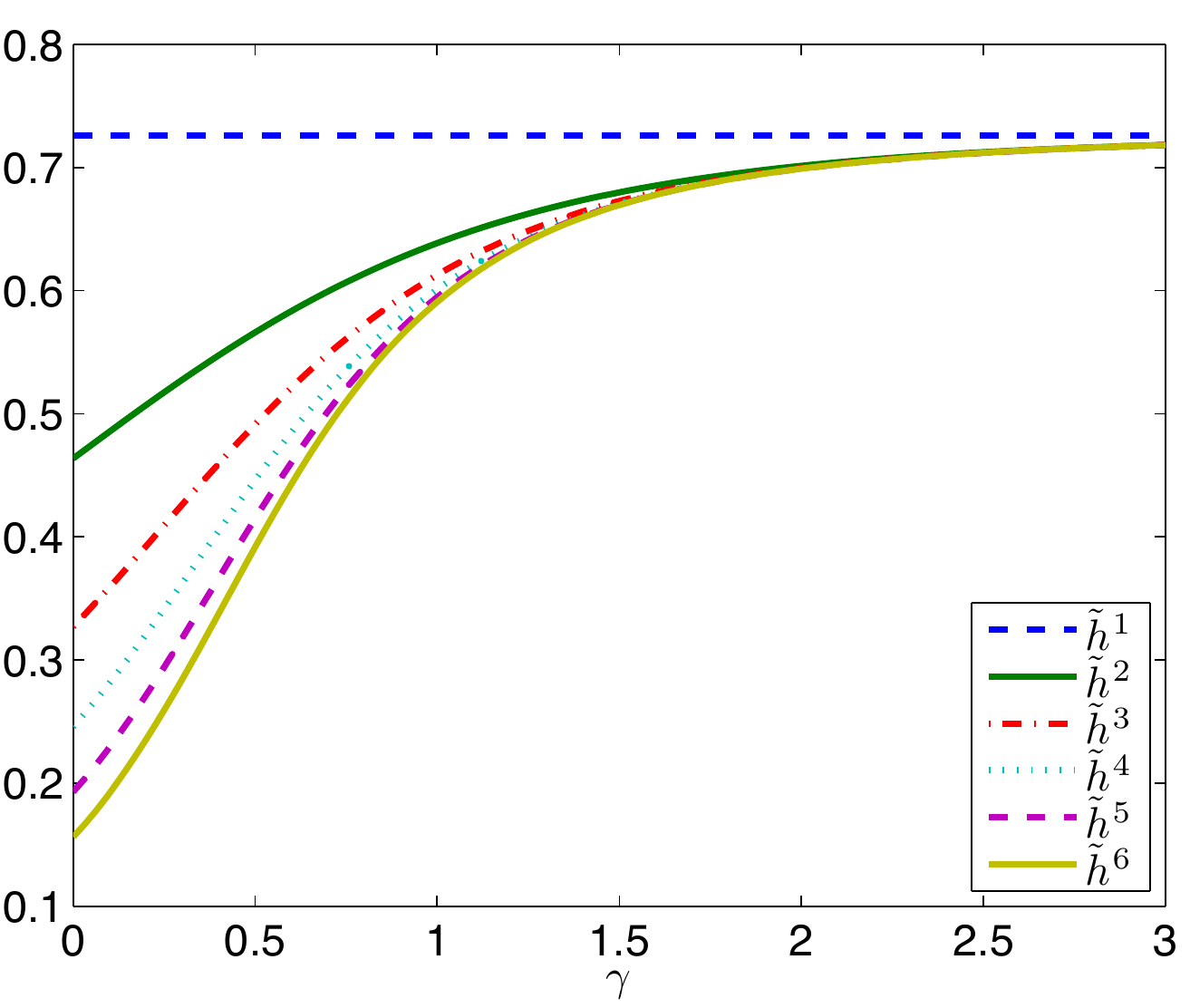}
  \caption{Two--parameter approximation
    \eqref{e:two_parameter_ansatz-new} of $H_{\rm Gauss}$ for
    different truncations. }
  \label{f:two_parameter_noalt}
\end{figure}

The sign alternating structure of the ansatz
\eqref{e:two_parameter_ansatz} is fundamental for the existence of the
critical point of $h(\gamma,A)$.  For the variational ansatz,
\begin{equation}
  \label{e:two_parameter_ansatz-new}
  a_p = A \abs{p}^{-\gamma} , \quad b_p =
  \tfrac{p^2}{3}, \quad p \in \Zo,
\end{equation}
we can redo the computations to obtain Figure
\ref{f:two_parameter_noalt}.  No critical point of $h(\gamma,A)$
exists for the sign-definite variational approximation
\eqref{e:two_parameter_ansatz-new}.

\section{Numerically Computed Gap Solitons}
\label{s:gap_solitons}

Using our observations from the Rayleigh-Ritz approximation, we are
motivated to solve the xNLS, \eqref{stat-NLS-system-epsilon}, directly
for existence of the gap solitons.  We note that in
\cite{Tasgal:2005p6335}, the authors explored the related problem
of broad band solitons of xNLCME truncated to two modes.

\subsection{Computation of the Gap Solitons}

We numerically solve equations \eqref{stat-NLS-system-epsilon} by
continuation.  Our starting point is the exact solution at $\eps=0$
\[
U_p(\zeta;\eps=0) = \tfrac{\sigma_p}{\sqrt{3 \Gamma}} \sech(p \zeta),
\]
where $\boldsymbol{\sigma}$ is a branch found in Section
\ref{s:gaussian_continuation} that led to a non-trivial solution at
$\eps=1$.  Iterating in $\eps$, we solve the system
\eqref{stat-NLS-system-epsilon} using \textsc{Matlab}'s {\tt bvp5c}
algorithm with absolute tolerance $10^{-4}$, relative tolerance
$10^{-8}$, on the domain $[0,25]$. {\tt bvp5c} is a nonlinear finite
difference algorithm for two-point boundary-value problems discussed
in \cite{shampine2003solving}.  We use the even
symmetry of the solutions to impose the boundary condition
$U_p'(0)=0$, and the artificial boundary condition
\[
U_p'(\zeta_{\max}) + p U_p(\zeta_{\max})=0.
\]

\begin{figure}
  \centering
  \subfigure{\includegraphics[width=2.5in]{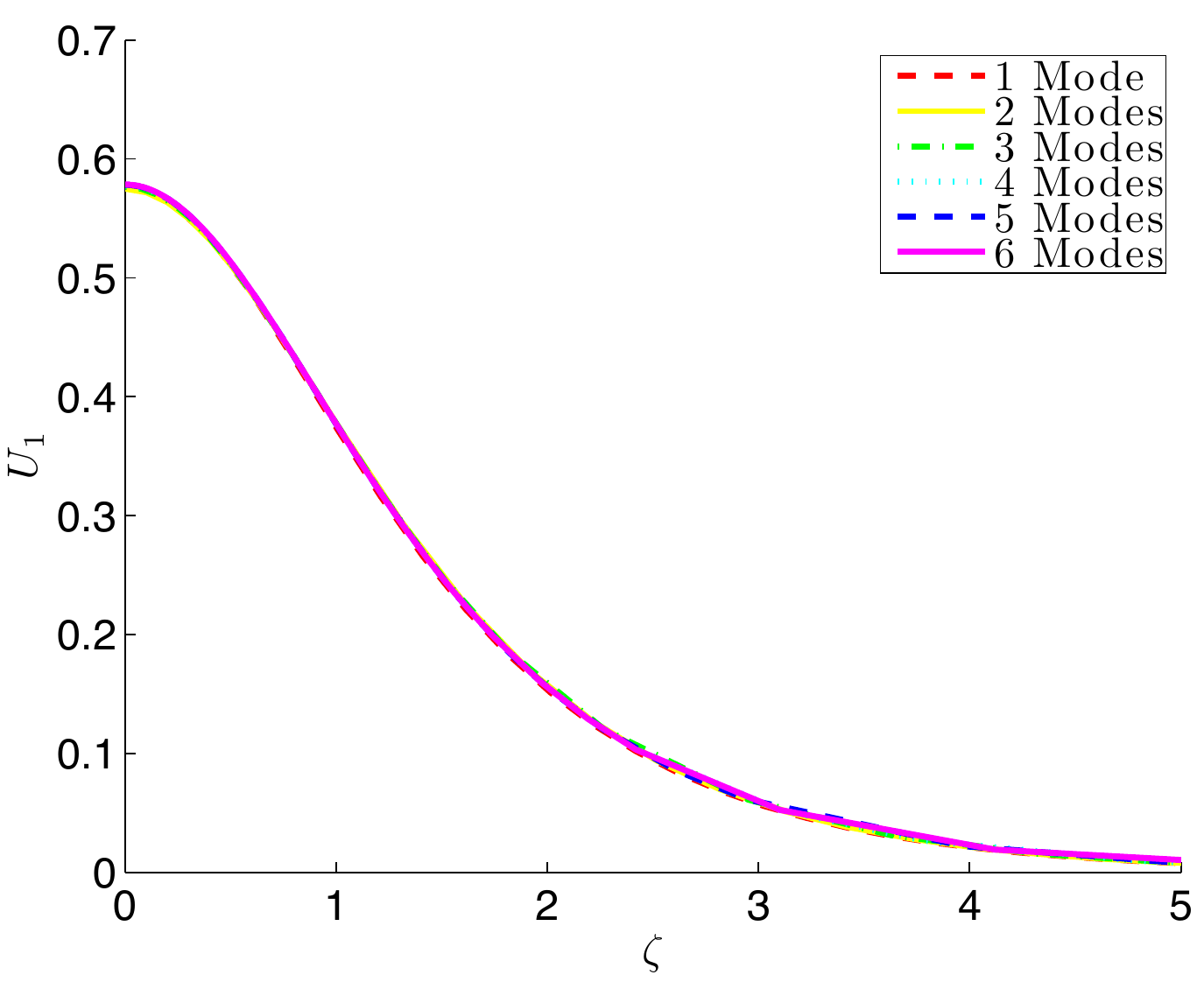}}
  \subfigure{\includegraphics[width=2.5in]{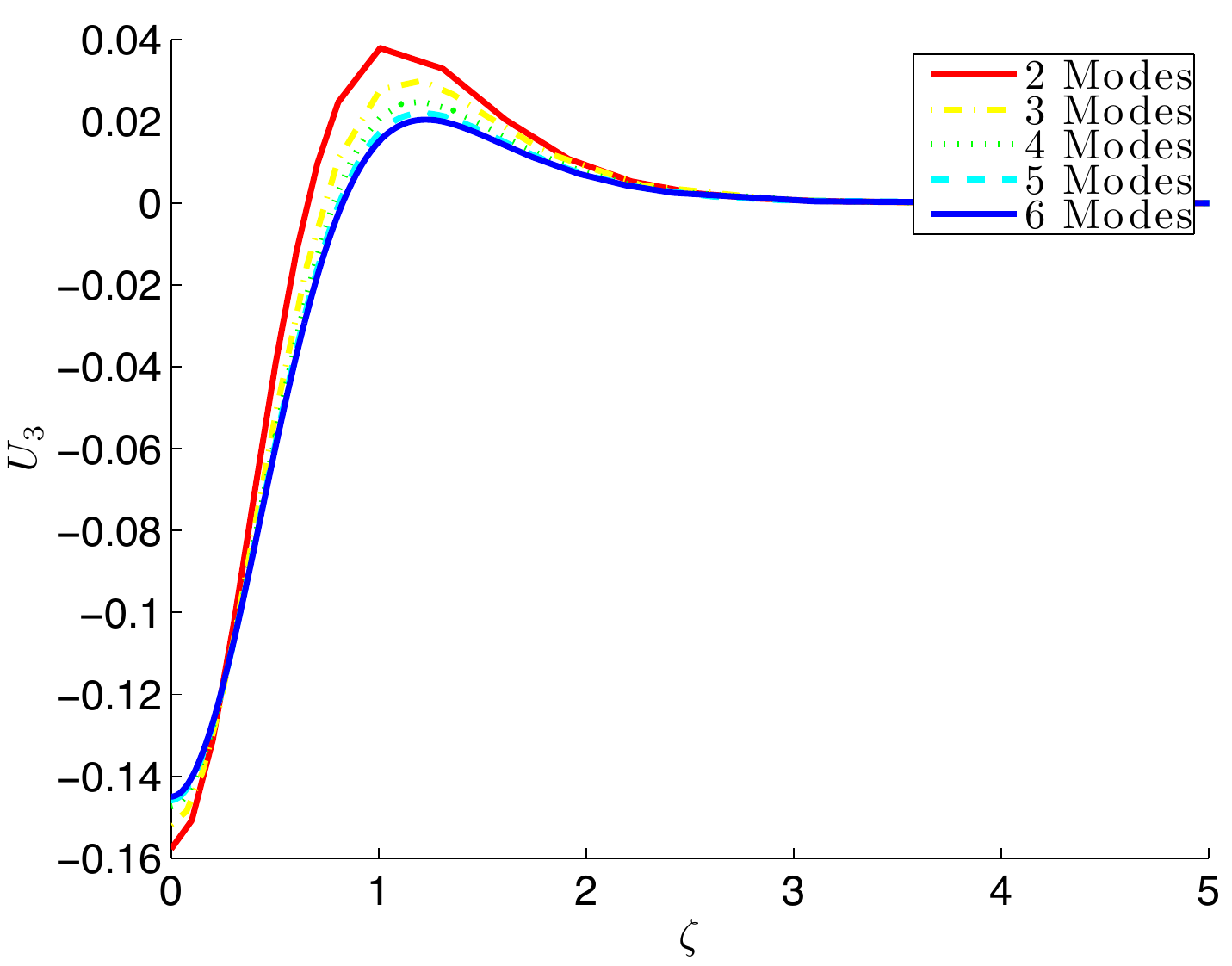}}
  \subfigure{\includegraphics[width=2.5in]{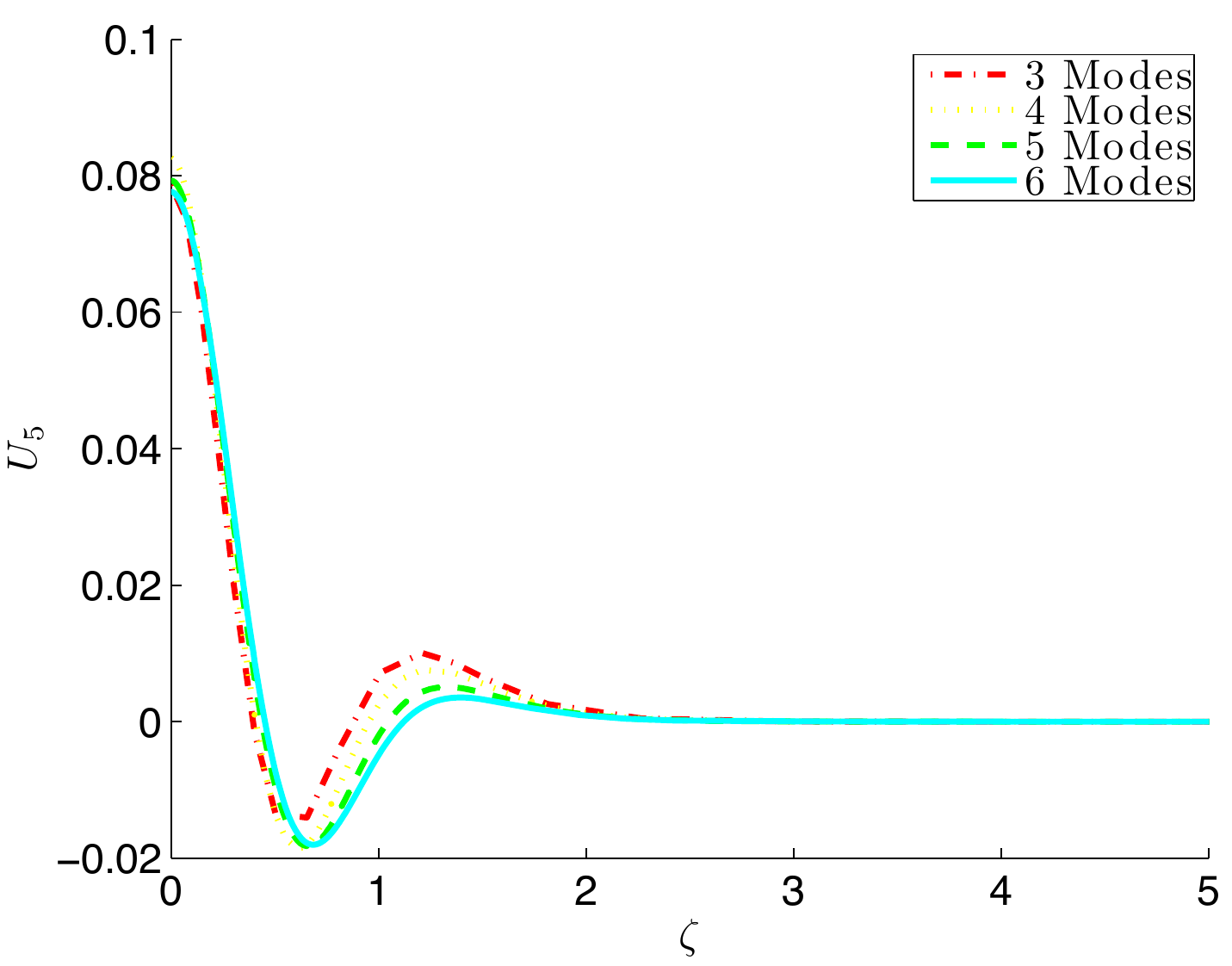}}
  \subfigure{\includegraphics[width=2.5in]{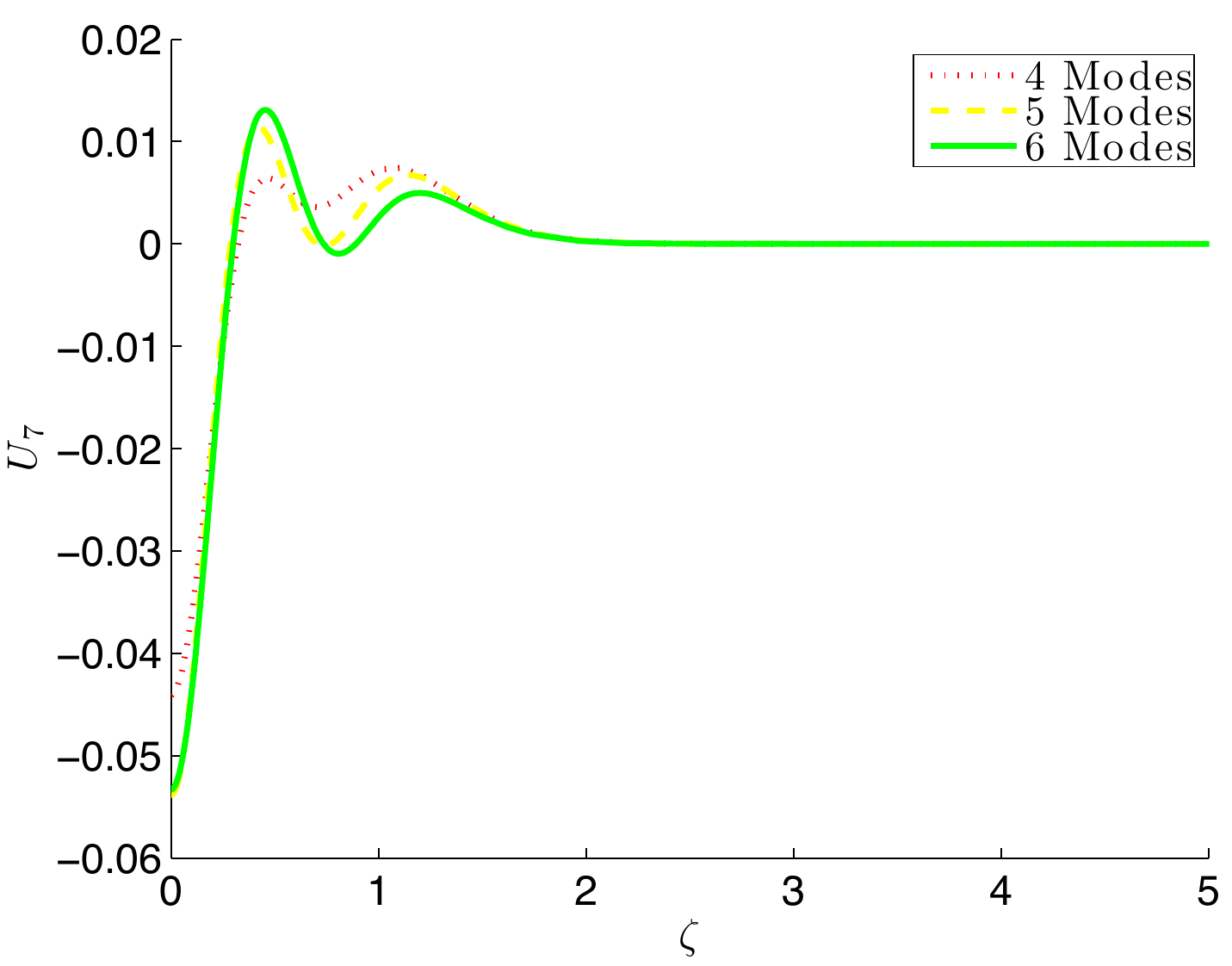}}
  \subfigure{\includegraphics[width=2.5in]{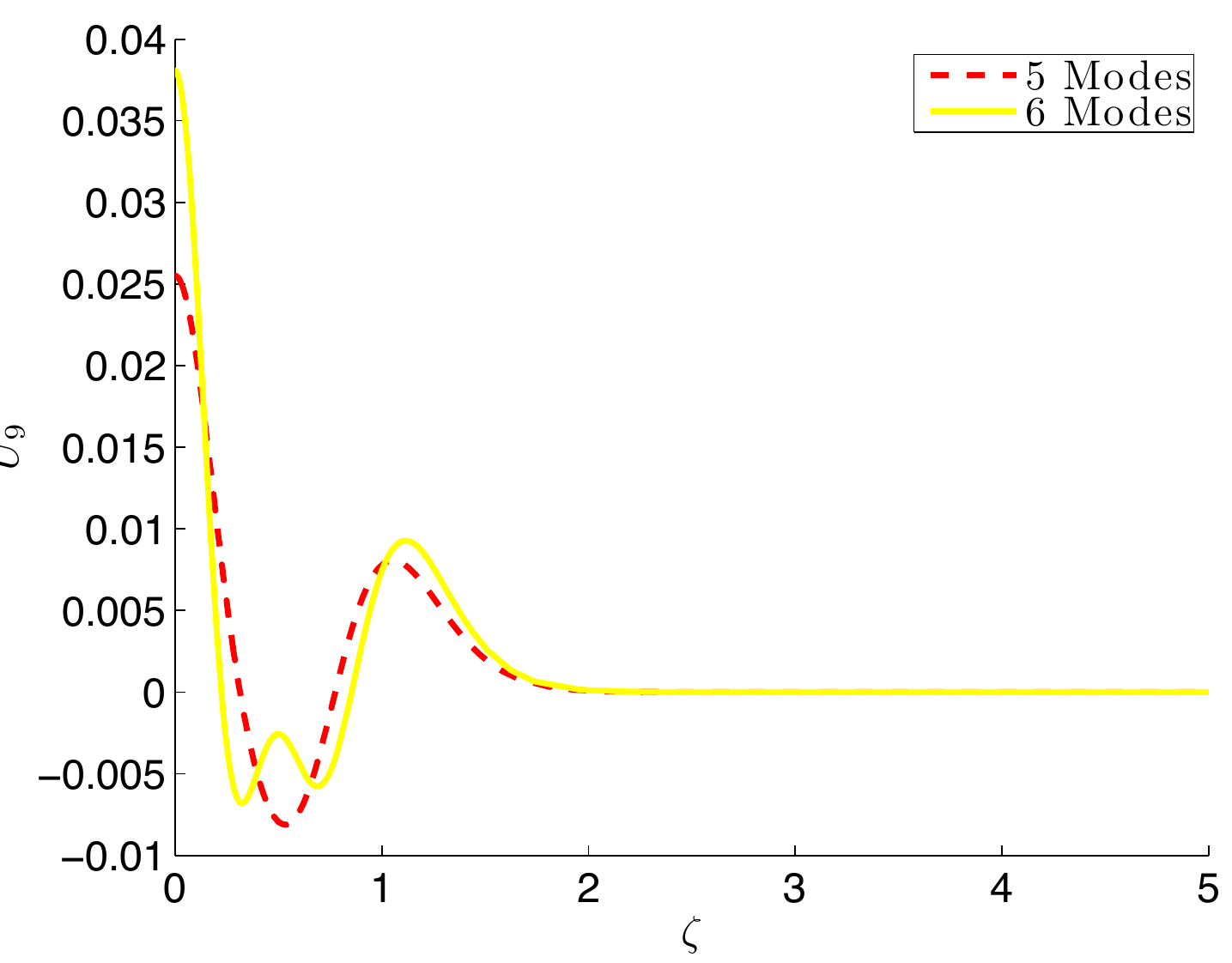}}
  \subfigure{\includegraphics[width=2.5in]{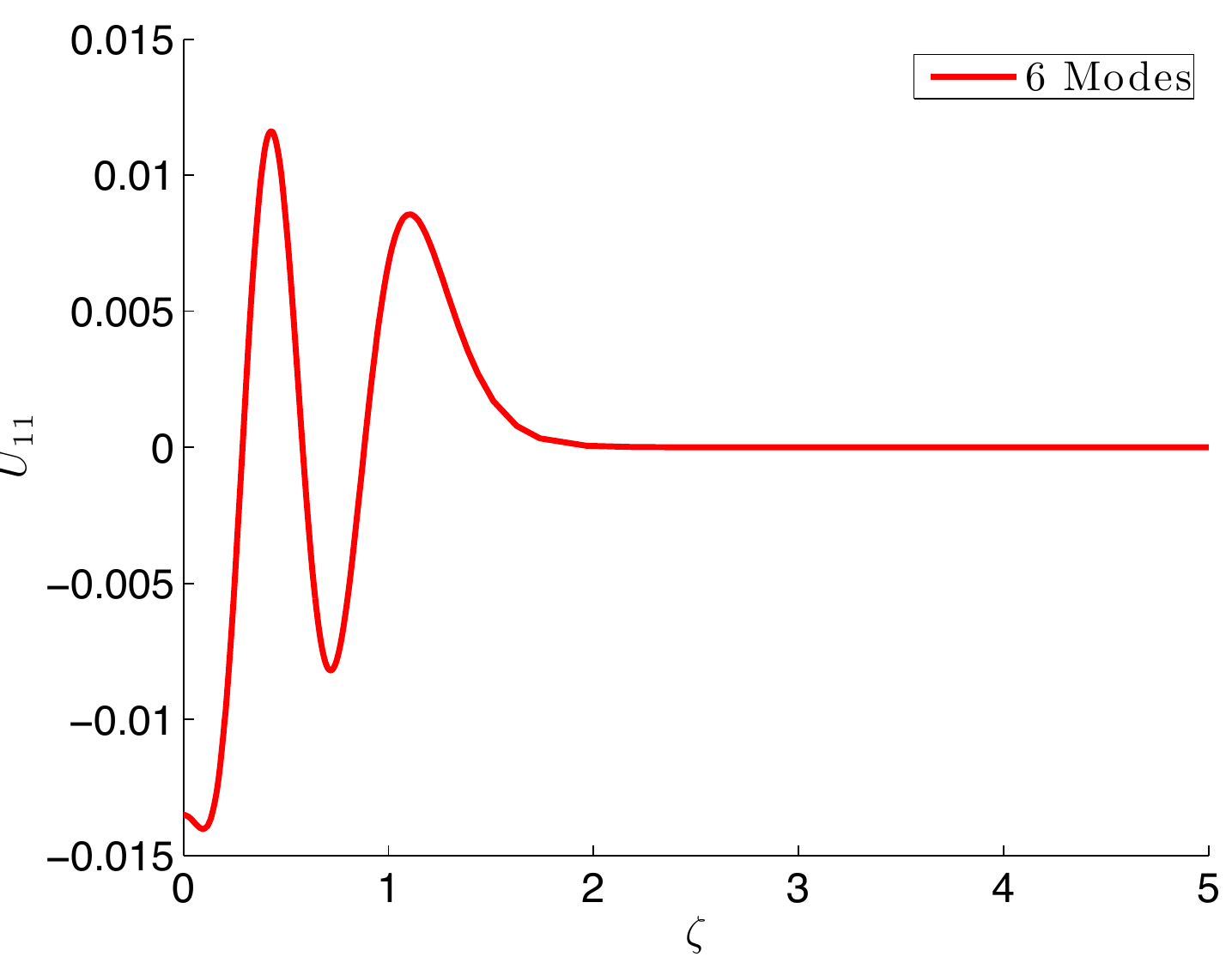}}
  \caption{Soliton profiles for the coupled NLS equations
    \eqref{stat-NLS-system}.}
  \label{f:soliton_profiles}
\end{figure}

\begin{figure}
  \centering \subfigure[Two Modes]{\includegraphics[width =
    3in]{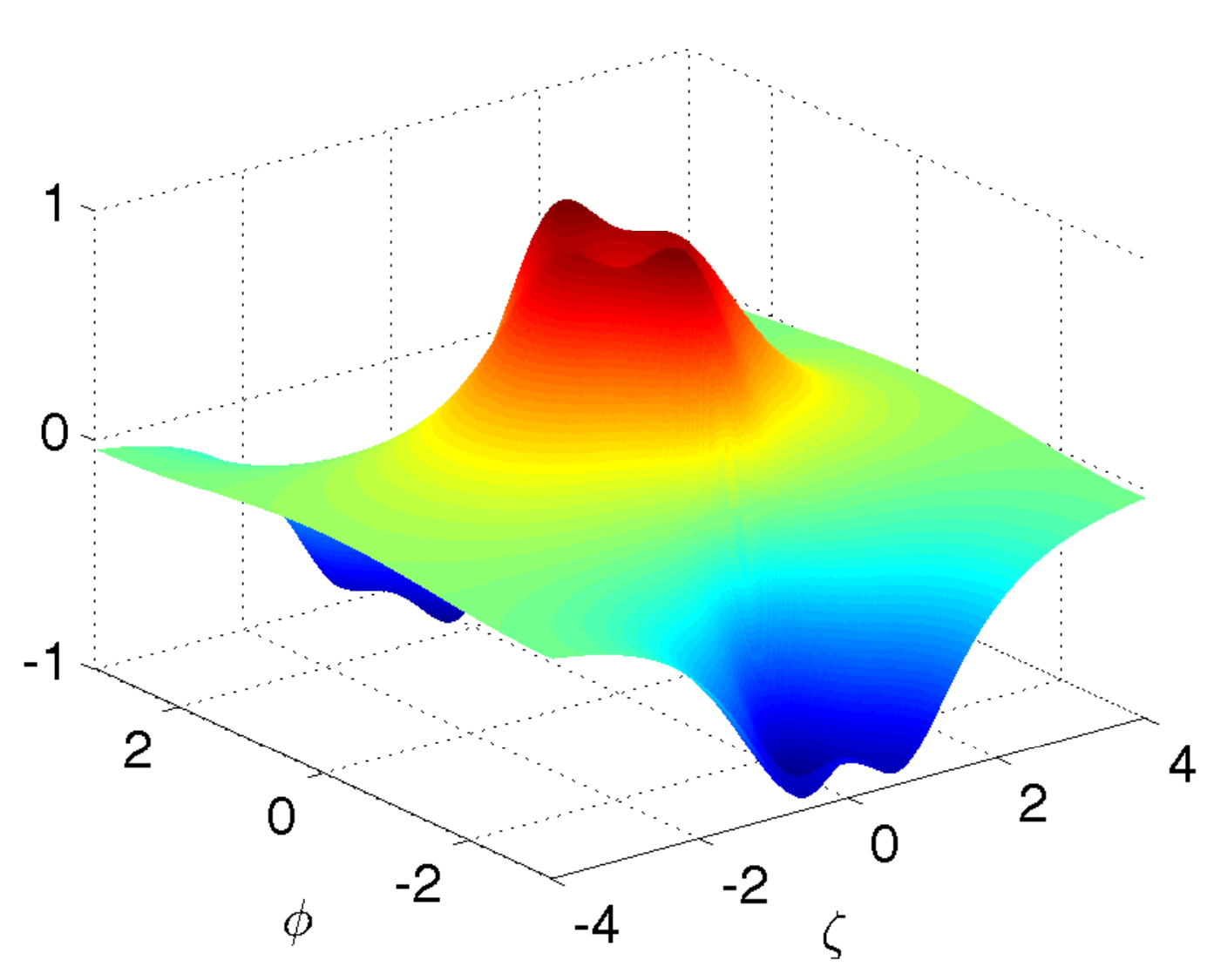}} \subfigure[Six
  Modes]{\includegraphics[width =
    3in]{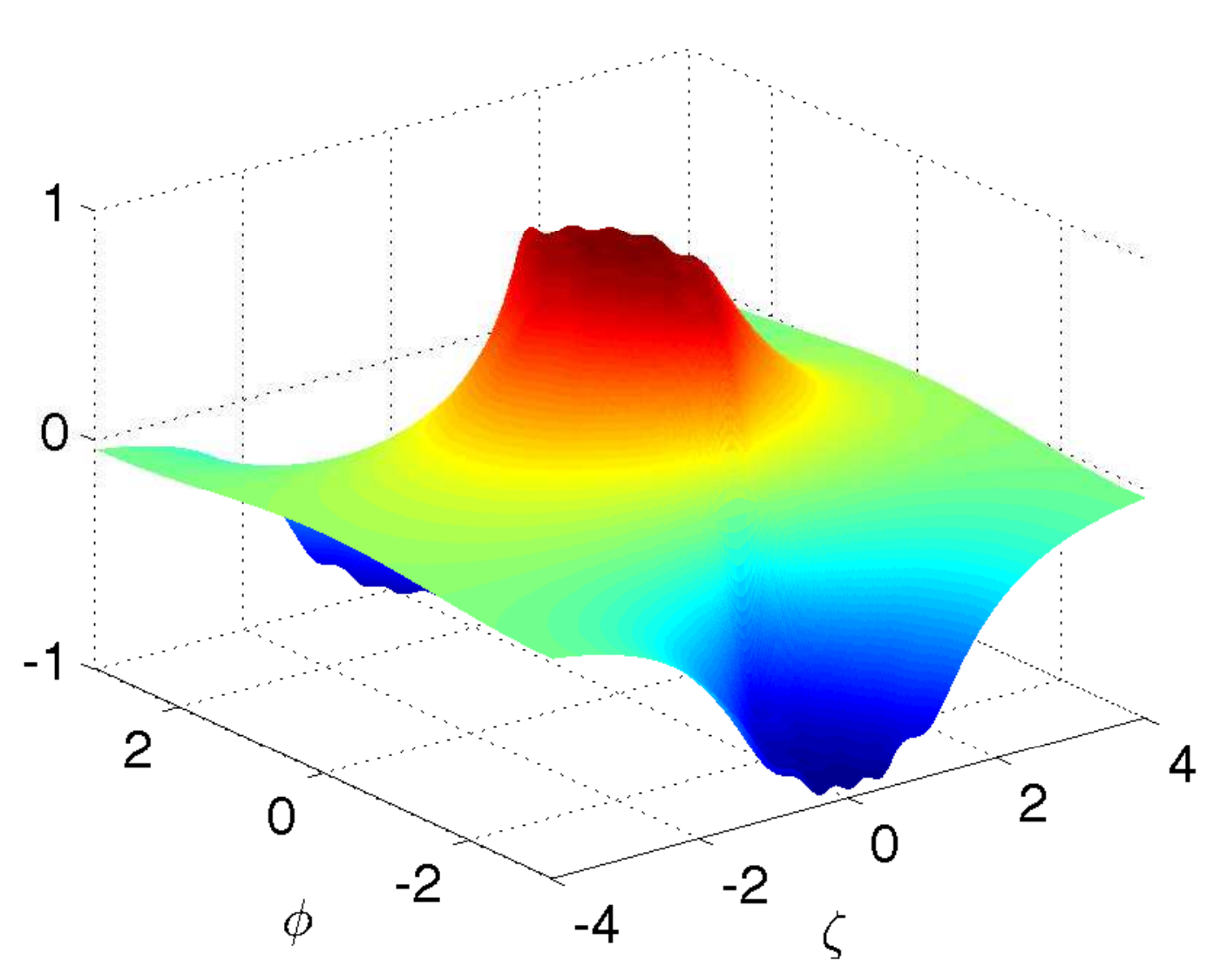}}
  \caption{The solution surface of the integral-differential equation
    \eqref{stat-NLS-system-fourier} generated by the truncated coupled
    NLS soliton on Figure \ref{f:soliton_profiles}.}
  \label{f:soliton_surface}
\end{figure}

The results for systems of up to six coupled NLS equations at
$\eps = 1$ appear in Figure \ref{f:soliton_profiles}.  As we can see
the amplitude decays in $p$, and they appear to approach some
asymptotic profile.  We conjecture that this profile persists as
additional modes are included.  Alternatively, the solution can be
expressed as $U(\zeta, \theta)$ by combining the Fourier modes. The
resulting solution surface of the integral-differential equation
\eqref{stat-NLS-system-fourier} appears in Figure
\ref{f:soliton_surface}.  The inclusion of additional harmonics
induces a more ornate structure near the extrema.

Though we have computed these finite truncation solutions, we
reiterate the question whether the corresponding solutions have finite
power. For our computed solutions, we find that the power, $N_{\rm
  xNLS}$, appears to converge and most of the power remains in the
first mode.  The data is given in Table \ref{t:soliton_l2}.

\begin{table}
  \centering
  \caption{Computed powers for the soliton profiles appearing in
    Figure \ref{f:soliton_profiles}.}
  \label{t:soliton_l2}
  \begin{tabular}{@{}l l l l@{}} \hline
    No. of Modes & $\norm{U_1}_{L^2}^2$ & $\frac{1}{2}N_{\rm xNLS}$ &
    $\frac{1}{2}N_{\rm xNLS}-\norm{U_1}_{L^2}^2$ \\
    \hline
    1 &0.66667  & 0.66667&0  \\
    2 &0.66982  & 0.68582&0.016000 \\
    3 & 0.67147 & 0.68929&0.017825\\
    4 &  0.67211 &0.69031&0.018201 \\
    5 &  0.67226&0.69070&0.018441\\
    6 & 0.67236 &0.69088&0.018523\\
    \hline
  \end{tabular}
\end{table}

\subsection{Gap Solitons in Time Dependent Simulations}

Small amplitude gap soliton solutions of the coupled NLS equation
\eqref{stat-NLS-system} can be used as initial conditions in the
coupled mode equations \eqref{e:mode_intro} to assess their stability
and robustness.  Once the solution $\{ U_p(\zeta) \}_{p \in \Zo}$ is
computed, the initial conditions for the time dependent simulation are
given by
\begin{equation}
  \label{e:soliton_ic}
  E^+_p(Z, 0)  = \mu U_p(\mu Z), \quad E^-_p(Z, 0)  = -\mu U_p(\mu Z),
  \quad p \in \Zo,
\end{equation}
with even modes set to zero.  By Theorem \ref{theorem-main-2}, the
small amplitude approximation is only accurate up to
$\bigo(\mu^2)$. We explore this small error as a source of the initial
perturbation.

We present the results of two and four mode systems.  In each case, we
truncated both the the system of coupled NLS equations
\eqref{stat-NLS-system} and the coupled mode equations
\eqref{e:mode_intro} at the same number of resolved modes. In our
simulations, we take as our constants
\[
v_g = 1, \quad N_0 = 0, \quad N_{2p} = 1, \quad \Gamma = 1.
\]
The simulations were performed with the indicated number of grid
points using a pseudo-spectral discretization and RK4 time stepping.
For both the two and four mode simulations, the initial conditions
\eqref{e:soliton_ic} were computed with greater precision than an in
the previous section; the absolute tolerance was $10^{-7}$ and the
relative tolerance was $10^{-9}$, and the domain was $[0,35]$.

\begin{figure}
  \centering
  \subfigure[]{\includegraphics[width=3in]{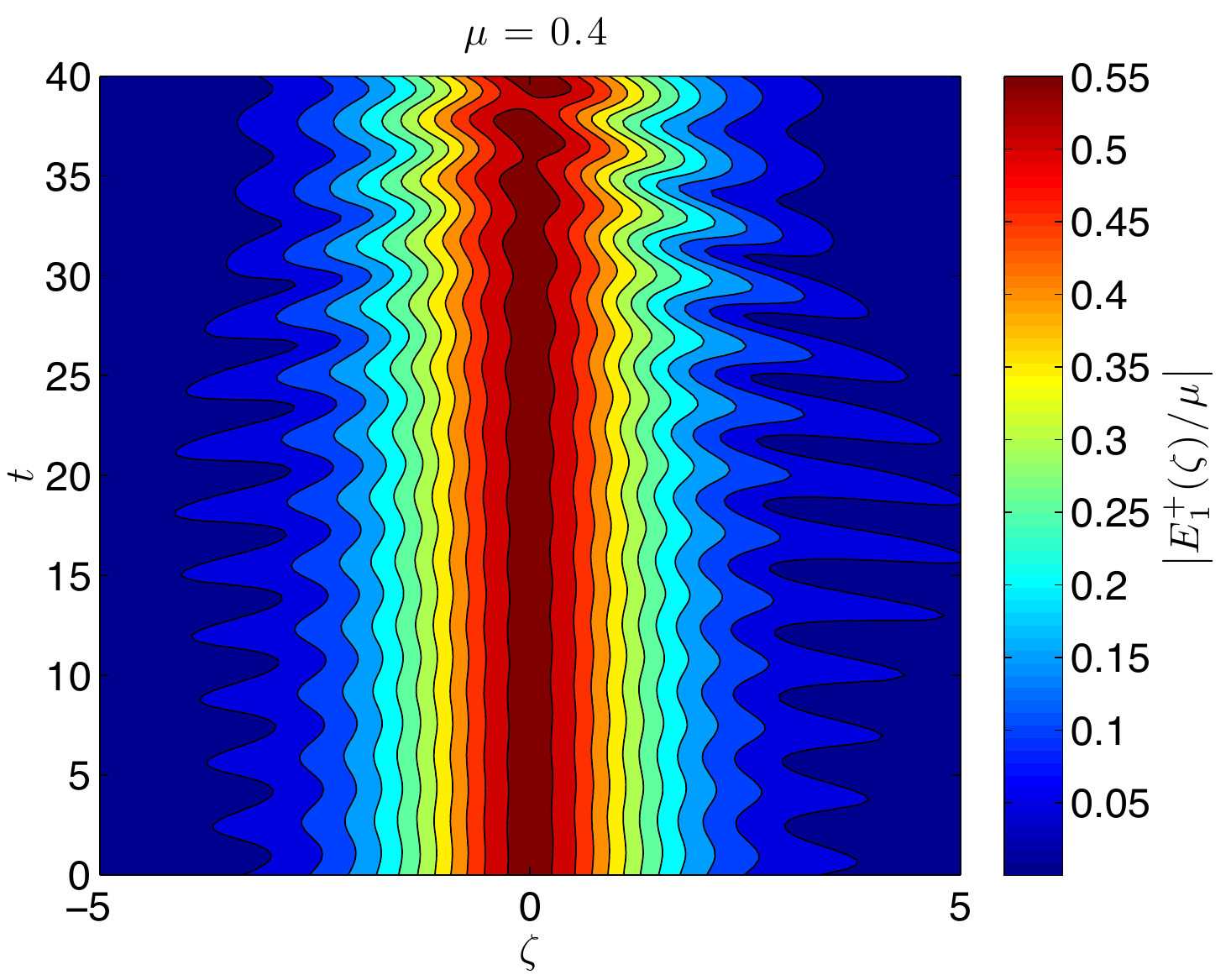}}
  \subfigure[]{\includegraphics[width=3in]{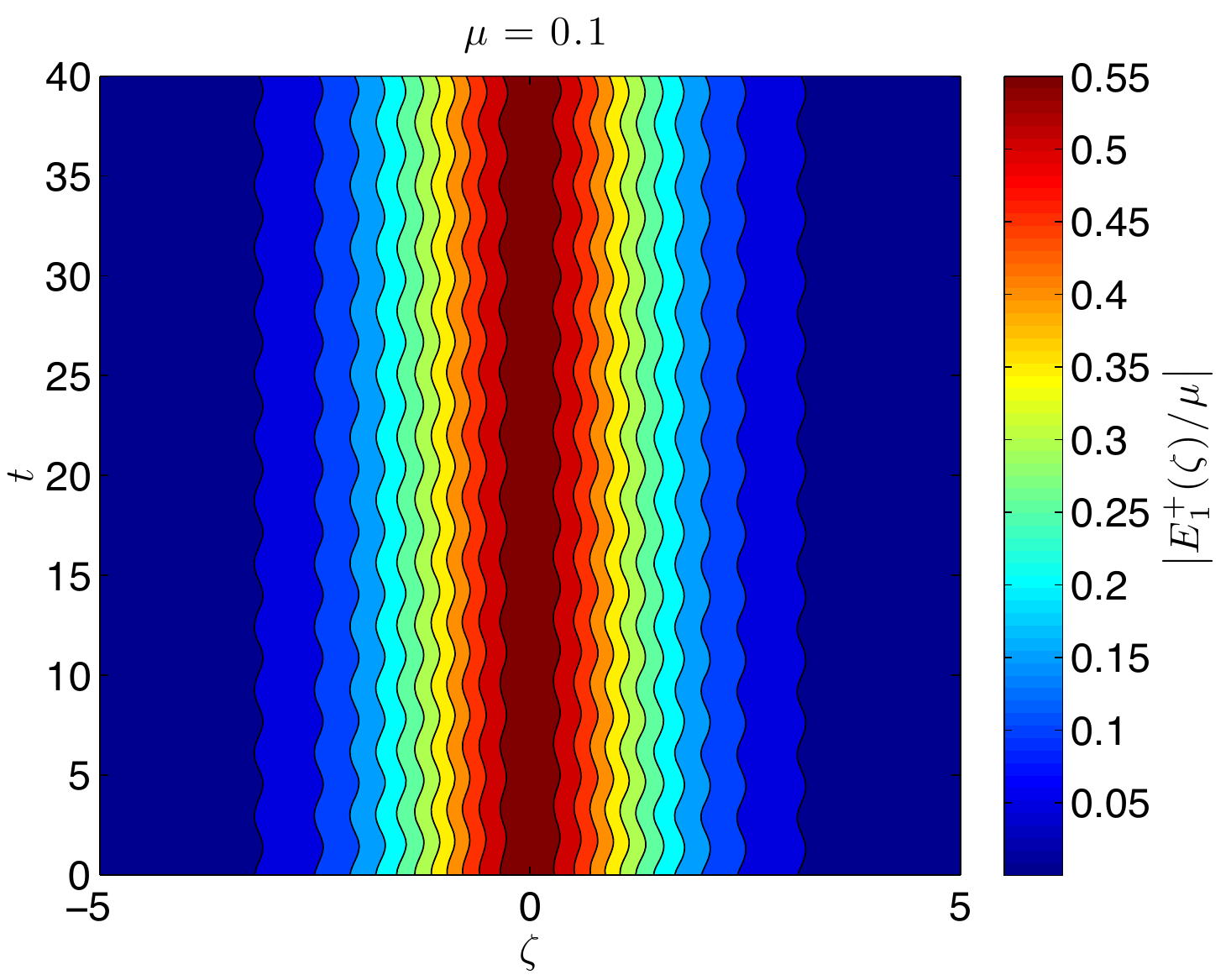}}\\
  \subfigure[]{\includegraphics[width=3in]{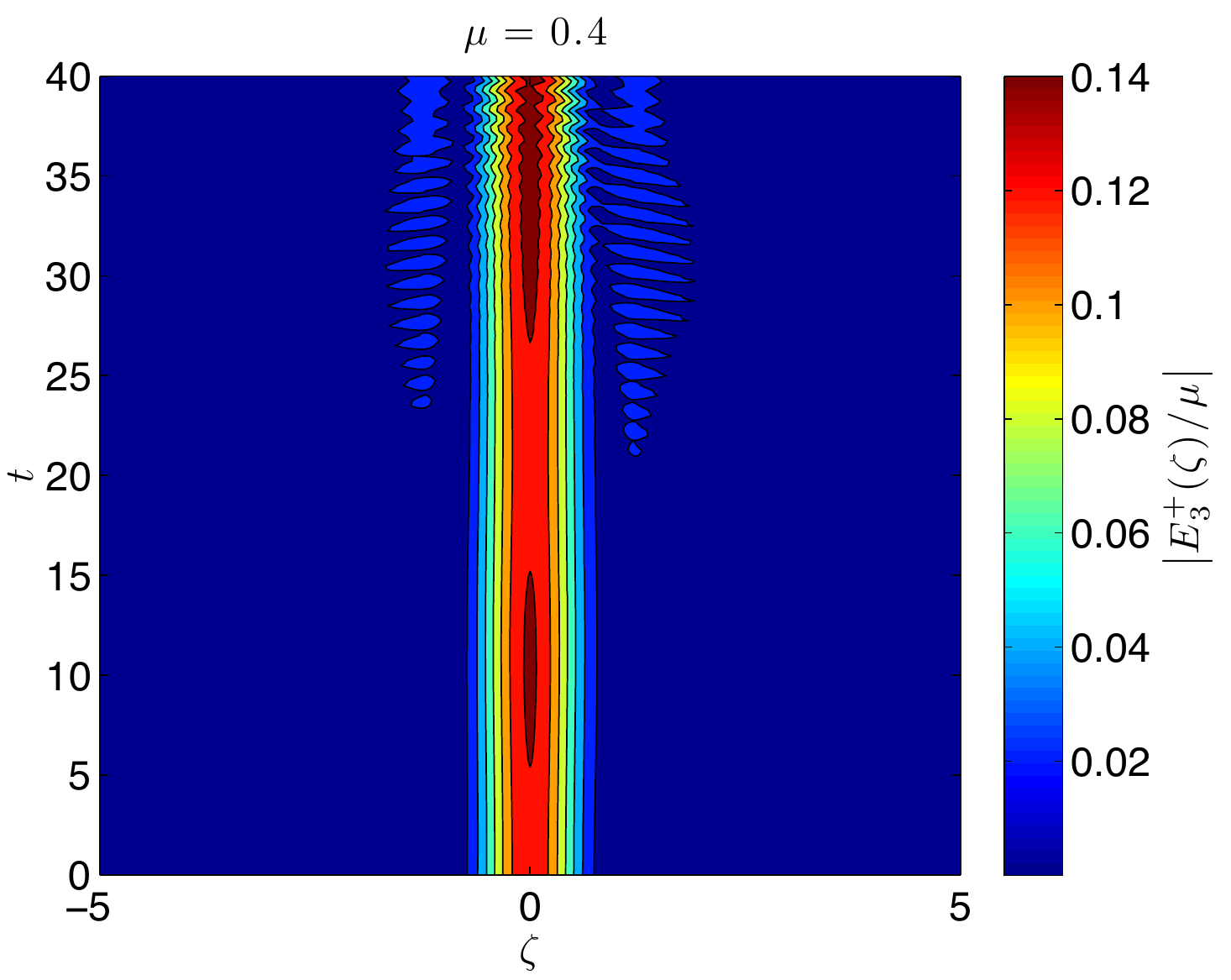}}
  \subfigure[]{\includegraphics[width=3in]{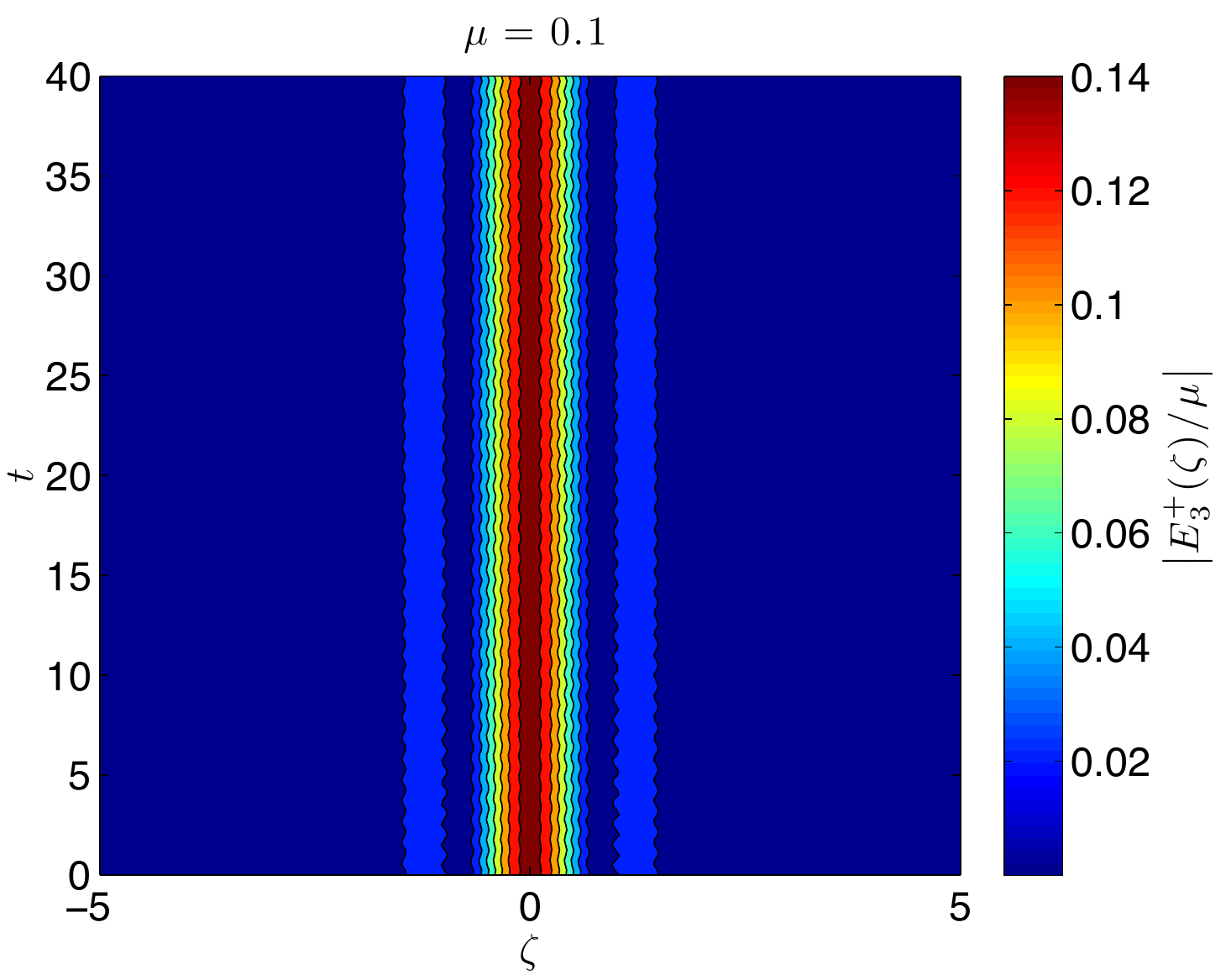}}
  \caption{Surfaces generated from simulations of the coupled mode
    system \eqref{e:mode_intro} truncated to four modes with initial
    data \eqref{e:soliton_ic}. The $\mu = .4$ simulations were
    computed on the domain $[-50, 50]$, and the $\mu=.1$ simulations
    were computed on the domain $[-200,200]$.  In both cases, there
    were 2048 grid points.}
  \label{f:4mode_surfs}
\end{figure}

In Figure \ref{f:4mode_surfs}, we plot the normalized time-space
surfaces of $\abs{E_p^+}$ from our simulations of the first four odd
modes.  For both values of $\mu$, the solution is persistent, but the
oscillations are greater for the larger value of $\mu$, and there is
some decoherence near the peak.  With the smaller value of $\mu$,
there is far less distortion.  Additional details of the dynamics are available
online in the following animations:
\begin{description}
\item[Two Mode Truncation] The following simulations were computed
  with 1024 grid points. The $\mu = .4$ simulations were computed on
    the domain $[-50, 50]$, the $\mu=.2$ simulations were computed on
    the domain $[-100,100]$, and the $\mu=.1$ simulations were
    computed on the domain $[-200,200]$.
{\small\begin{itemize}
\item \url{http://www.math.toronto.edu/simpson/files/media/broadband/mode1_13.mp4}
\item \url{http://www.math.toronto.edu/simpson/files/media/broadband/mode3_13.mp4}
\end{itemize}}

\item[Four Mode Truncation]  The following simulations were computed
  with 2048 grid points. The $\mu = .4$ simulations were computed on
    the domain $[-50, 50]$, the $\mu=.2$ simulations were computed on
    the domain $[-100,100]$, and the $\mu=.1$ simulations were
    computed on the domain $[-200,200]$. 
{\small\begin{itemize}
\item \url{http://www.math.toronto.edu/simpson/files/media/broadband/mode1_1357.mp4}
\item \url{http://www.math.toronto.edu/simpson/files/media/broadband/mode3_1357.mp4}
\item \url{http://www.math.toronto.edu/simpson/files/media/broadband/mode5_1357.mp4}
\item \url{http://www.math.toronto.edu/simpson/files/media/broadband/mode7_1357.mp4}
\end{itemize}}
\end{description}
As one would expect, there is better agreement between the approximate small
amplitude soliton and the time dependent simulation as $\mu \to 0$.
However, for all values of $\mu$ presented, there is a persistence of
the localization, even if there is distortion to some of the fine
structure in the higher harmonics.  All of this suggests the gap
solutions are robust.

\begin{figure}
  \centering
  \subfigure[]{\includegraphics[width=3in]{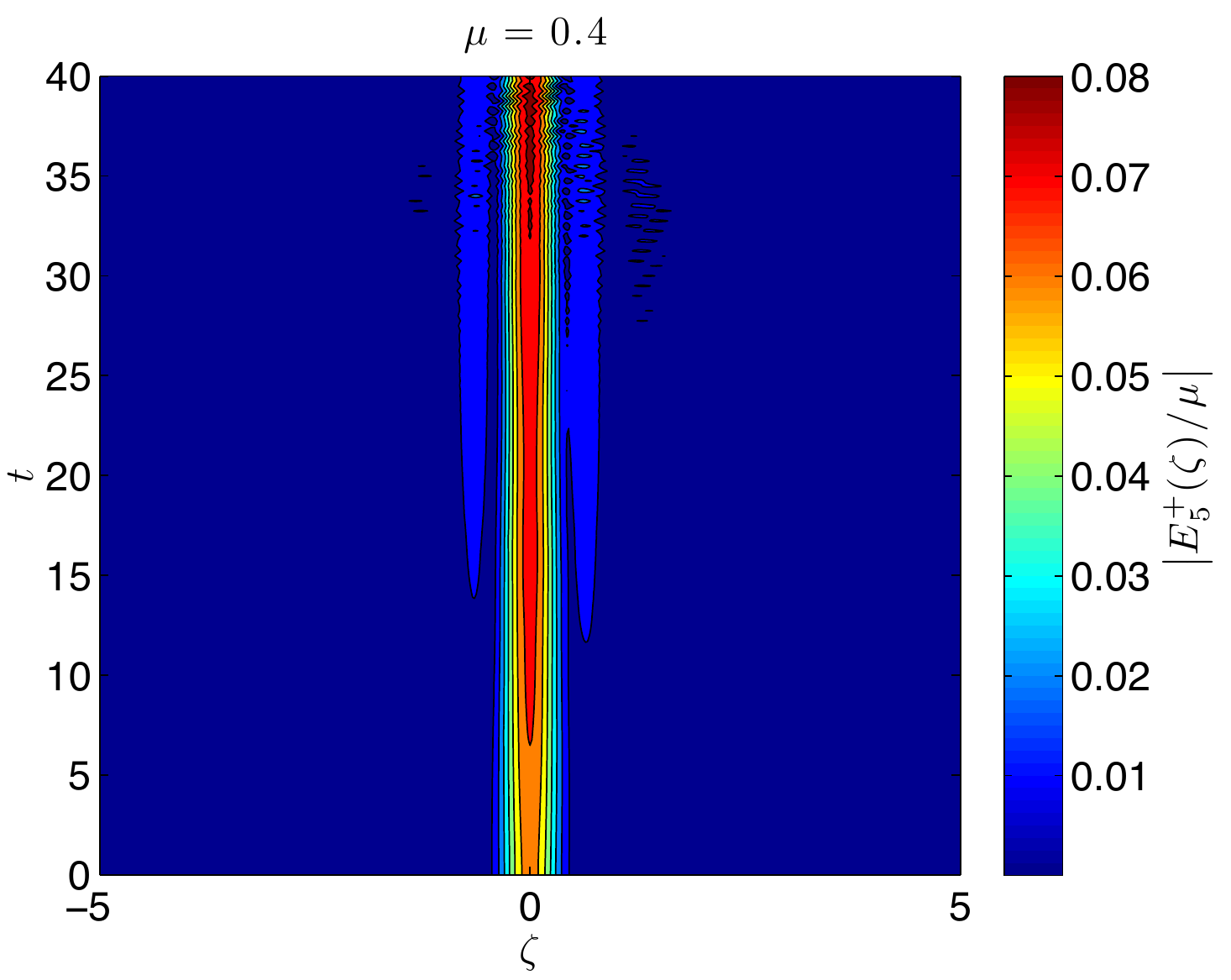}}
  \subfigure[]{\includegraphics[width=3in]{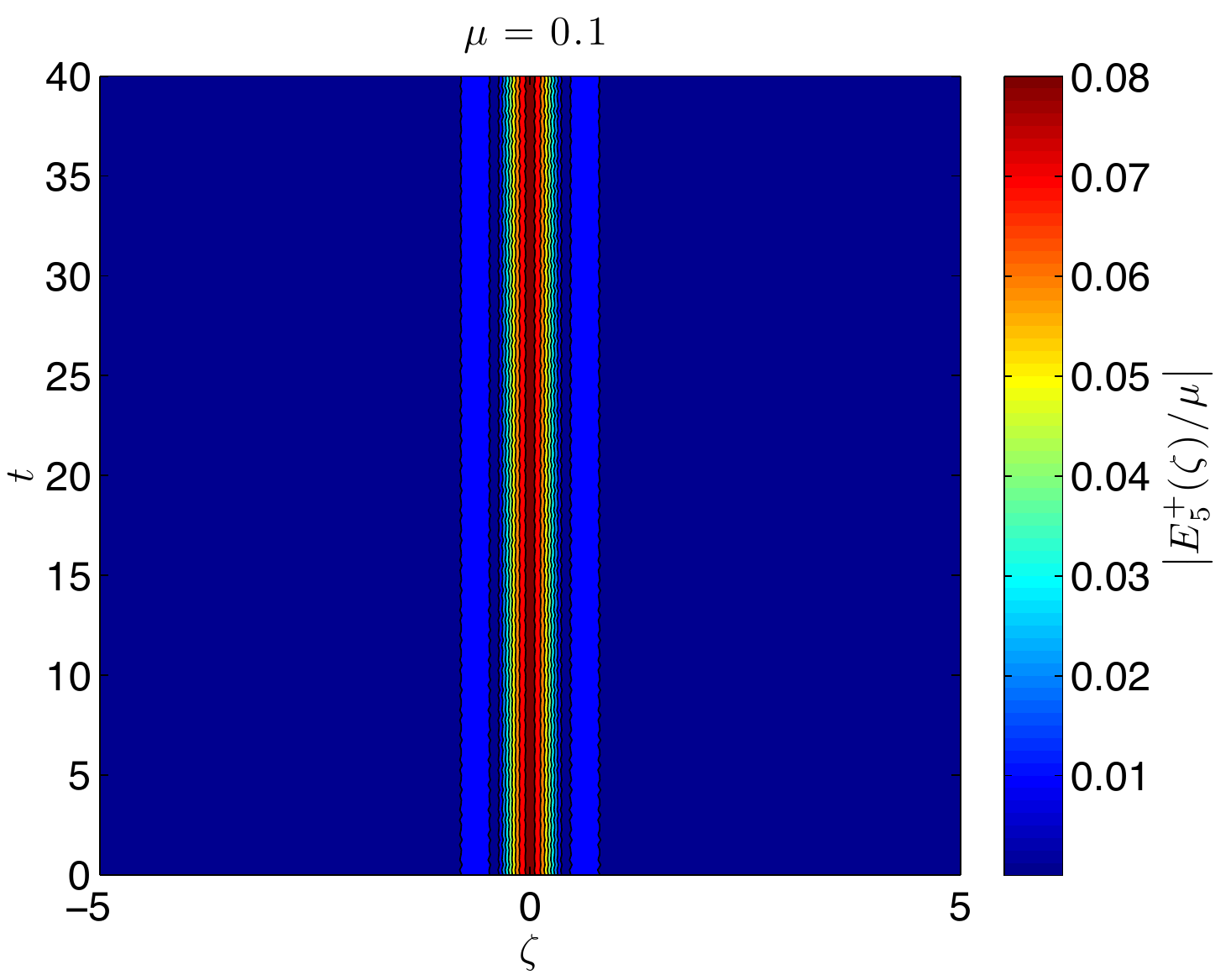}}\\
  \subfigure[]{\includegraphics[width=3in]{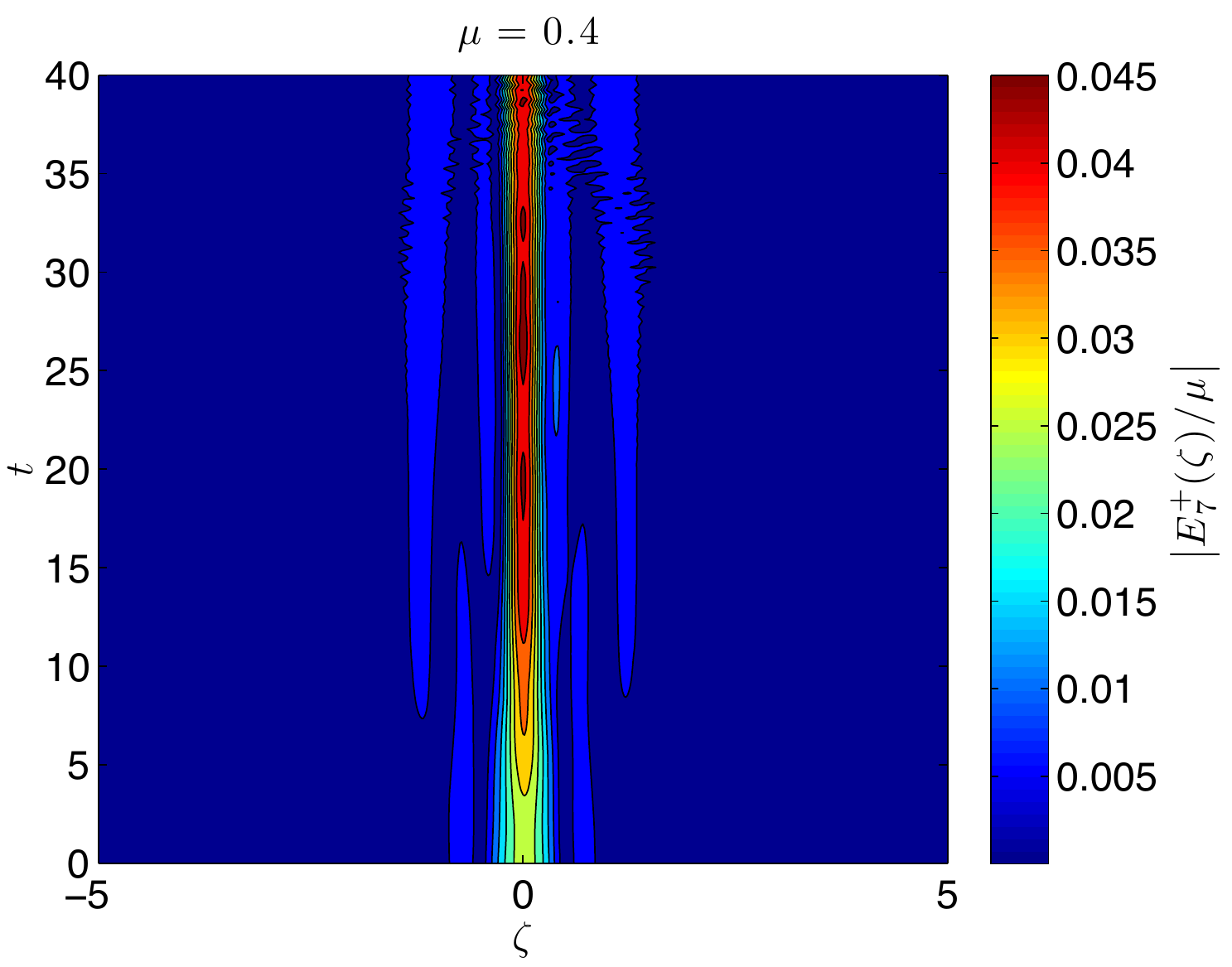}}
  \subfigure[]{\includegraphics[width=3in]{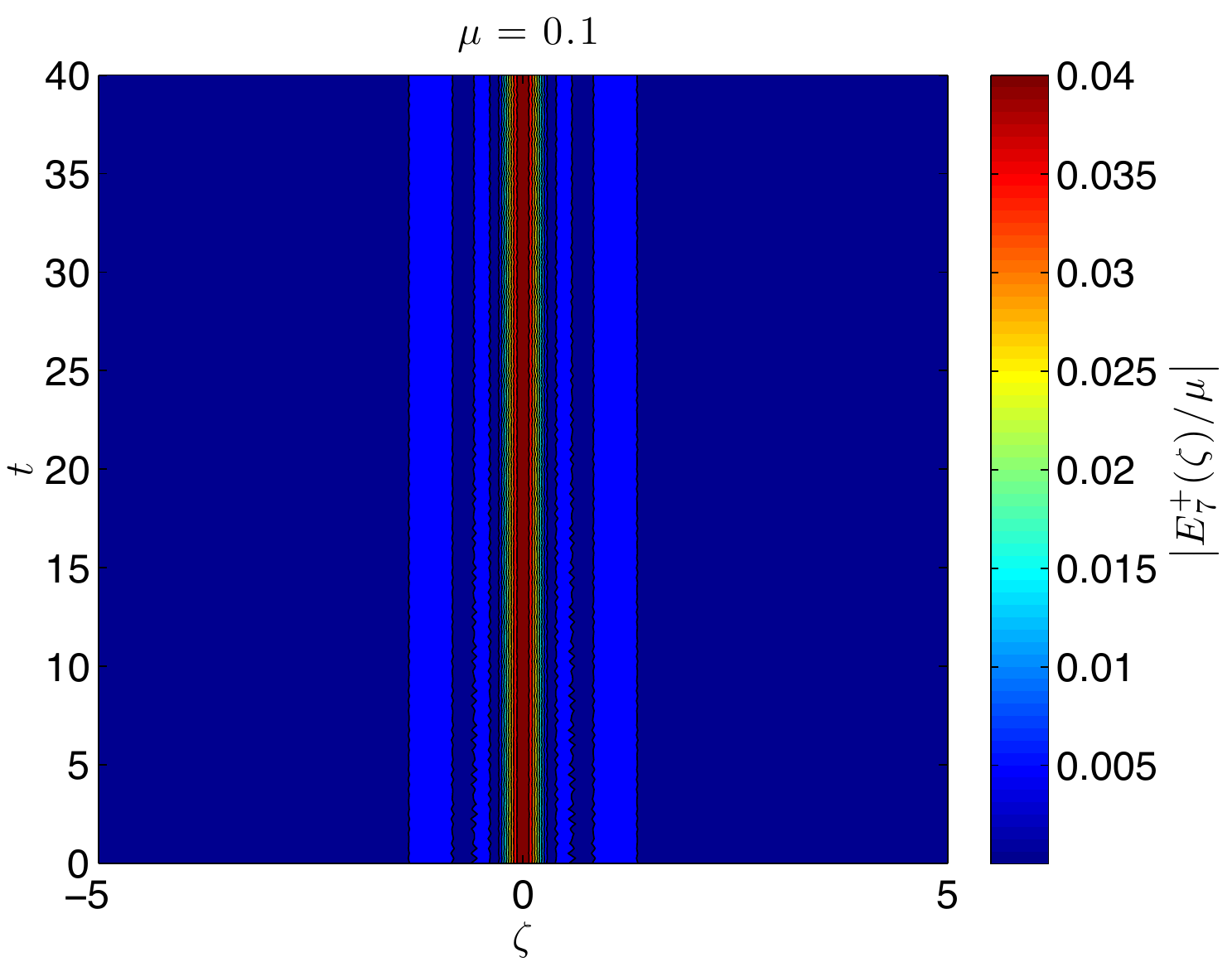}}
  \caption*{Figure \ref{f:4mode_surfs} continued.}
  % \label{f:4mode_surfs}
\end{figure}

Many other experiments are possible; simulating with more modes,
simulating with larger values of $\mu$, and seeding the initial
conditions of a smaller system into a larger system.  In the previous
work \cite{simpson2010}, the exact gap soliton \eqref{gap-soliton} was
used as an initial condition for successively larger truncations of
the extended coupled mode system \eqref{e:mode_intro}.

\section{Open Problems}
\label{s:open}

We conclude this work with a discussion of open problems concerning
the existence of non-trivial localized solutions of xNLCME and xNLS,
arising for the case of a refractive index composed of an infinite
array of Dirac delta functions. Some of the challenges include:
\begin{itemize}
\item Prove the existence of a non-trivial critical point to $h$,
  \eqref{approximation-h}, the single parameter Rayleigh-Ritz
  approximation,
\item Prove the existence of a non-trivial solution to $H_{\rm
    Gauss}$, \eqref{hamiltonian-ansatz}, the Gaussian Rayleigh-Ritz
  approximation,
\item Prove the existence of a non-trivial solution to the coupled NLS
  equations \eqref{stat-NLS-system},
\item Prove the existence of a non-trivial solution to the coupled
  mode equations \eqref{stat-cme-system}.
\end{itemize}
By ``non-trivial'', we mean a solution in which all modes are
active and non-zero.  It would also be of interest to obtain proofs
of existence for arbitrarily large finite truncations of these
problems.  Intimately connected with the last two challenges is the
question of appropriate function spaces. As discussed in Section
\ref{s:tails}, our variational approximations live in the function
space $X^s$ for $1 < s < \gamma \approx 1.26$ for which our Theorems
\ref{theorem-main-2}  and\ref{thm:xnls_mono} are stated. The upper
value on $s$ that ensures that the interval is nonempty is only
approximated numerically from the ``rough'' variational
approximation.   Of course, it is also possible that such solutions may not exist.  A
counterexample would also be of interest.

Modeling the nonlinear Maxwell equation with refractive
index given by a periodic sequence of Dirac delta-functions is a
challenging problem both analytically and numerically. Results of our
work give a starting point to further exploration of this system, and
the evolution of its localized excitations.  The question of localized
solutions for xNLCME for less restrictive, and more physical,
refractive indices is also of great interest.  

%\clearpage

% \bibliographystyle{abbrv} \bibliography{broadband}

\end{document}

%% file: gsmacros.tex
\newcommand{\abs}[1]{\lvert#1\rvert}
\newcommand{\norm}[1]{\lVert#1\rVert}
\newcommand{\paren}[1]{\left(#1\right)}
\newcommand{\bracket}[1]{\left[#1\right]}
\newcommand{\set}[1]{\left\{#1\right\}}

\newcommand{\bigo}{\mathcal{O}}

\newcommand{\eps}{\epsilon}

\newcommand{\dZ}{\partial_{Z} }
\newcommand{\dT}{\partial_{T} }
\newcommand{\Ep}{E^+ }
\newcommand{\Em}{E^-}

\newcommand{\cc}{{\mathrm{c.c.}}}

\newcommand{\im}{\mathrm{i}}

\newcommand{\nn}{\nonumber}

\newcommand{\calN}{\mathcal{N}}
\newcommand{\N}{\mathbb{N}}

\newcommand{\R}{\mathbb{R}}
\newcommand{\Z}{\mathbb{Z}}
\newcommand{\Zodd}{\mathbb{Z}_{\rm odd}}
\newcommand{\T}{\mathbb{T}}
\newcommand{\Zo}{\mathbb{Z}_{\mathrm{odd}}}

\newcommand{\sech}{\mathrm{sech}}
\newcommand{\sign}{\mathrm{sign}}

\newtheorem{theorem}{Theorem}
\newtheorem{prop}{Proposition}[section]

\newtheorem{conj}[prop]{Conjecture}